\newcommand*{\ATLASLATEXPATH}{latex/}
\documentclass[UKenglish,texlive=2016,cernpreprint]{\ATLASLATEXPATH atlasdoc}
\usepackage[subfigure]{\ATLASLATEXPATH atlaspackage}
\usepackage{\ATLASLATEXPATH atlasbiblatex}
\usepackage{multirow}
\usepackage{bigstrut}
\usepackage{longtable}

\usepackage{\ATLASLATEXPATH atlascontribute}

\usepackage{\ATLASLATEXPATH atlasphysics}
\graphicspath{{logos/}{figures/}}
\usepackage{Moriond2017_HZZ_HighMass-defs}
\AtlasTitle{Search for heavy $ZZ$ resonances in the \llll and \llvv final states using proton--proton collisions  at $\sqrt{s}= 13$~\tev\ with the ATLAS detector}
\PreprintIdNumber{CERN-EP-2017-251}
\AtlasJournalRef{Eur. Phys. J. C 78 (2018) 293}
\AtlasDOI{DOI:10.1140/epjc/s10052-018-5686-3}
\AtlasAbstract{%
A  search for heavy resonances decaying into a pair of \Z bosons leading to \llll and  \llvv final states, where $\ell$ stands for either an electron or a muon, is presented.  The search uses proton--proton collision data at a centre-of-mass energy of 13~\tev\ corresponding to an integrated luminosity of \lumi\ collected with the ATLAS detector during 2015 and 2016 at the Large Hadron Collider. Different mass ranges for the hypothetical resonances are considered, depending on the final state and model. The different ranges span between 200~\gev\ and 2000~\gev.  The results are interpreted as  upper limits on the production cross section of a spin-0 or spin-2 resonance. 
The  upper limits for the spin-0 resonance are translated to exclusion contours in the context of Type-I and Type-II two-Higgs-doublet models, while those for the spin-2 resonance are used to constrain the Randall--Sundrum model with an extra dimension giving rise to spin-2 graviton excitations.
}

\hypersetup{pdftitle={ATLAS document},pdfauthor={The ATLAS Collaboration}}
\begin{document}
\maketitle
\tableofcontents
\clearpage
\section{Introduction}
\label{sec:intro}
In 2012, the ATLAS and CMS Collaborations at the LHC discovered a new
particle~\cite{HIGG-2012-27,CMS-HIG-12-028}, an important milestone in the understanding of the
mechanism of electroweak (EW) symmetry breaking~\cite{Englert:1964et,Higgs:1964pj,Guralnik:1964eu}.
Both experiments have confirmed that the spin, parity and couplings of the new particle are
consistent with those predicted for the Standard Model (SM) Higgs
boson~\cite{HIGG-2014-06,CMS-HIG-14-009,HIGG-2013-01} (denoted by \SMH throughout this paper). They 
measured its mass to be $\msmallh = 125.09 \pm 0.21 \mathrm{(stat)} \pm
0.11 \mathrm{(syst)}$~\gev~\cite{HIGG-2014-14} and reported recently on a combination of
measurements of its couplings to other SM particles~\cite{HIGG-2015-07}.
 
One important question is whether the newly discovered particle is part of an extended scalar sector as postulated by various extensions to the Standard Model such as the two-Higgs-doublet model
(2HDM)~\cite{Branco:2011iw}. These extensions predict additional Higgs~bosons, motivating searches in an extended range of mass. 

This paper reports on two searches for a heavy resonance
decaying into two SM $Z$ bosons, encompassing the final states $\zztollll$ and $\zztollvv$ where $\ell$ 
stands for either an electron or a muon and $\nu$ stands for all three neutrino flavours.  These final states are referred to as $\llll$ and $\llvv$ respectively.

It is assumed that an additional Higgs boson (denoted as \res throughout this paper) would be produced predominantly via gluon--gluon fusion
(ggF) and vector-boson fusion (VBF) processes, but that the ratio of the two production mechanisms is
unknown in the absence of a specific model. 
For this reason, the results are interpreted separately for the ggF and VBF production modes, with events being classified into ggF- and VBF-enriched categories in both final states, as discussed in Sections~\ref{sec:h4lanal} and~\ref{sec:hllvvanal}.  With good mass resolution and a high signal-to-background ratio, the \llll\ final state is well suited to a search for a narrow resonance with mass $\mres$ between $200~\gev$  and $1200~\gev$. The \llvv\ search covers the $300~\gev<\mres<1400~\gev$ range and dominates at high masses due to its larger branching ratio. 
  
These searches look for an excess in distributions of the four-lepton invariant mass, \mfl, for the \llll final state,
and the transverse invariant mass, \mt, for the \llvv final state,
as the escaping neutrinos do not allow the full reconstruction of the final state. The transverse invariant mass is defined as:
\begin{equation*}
  \mt \equiv \sqrt{ \left[ {\sqrt{\mz^{2} + \left(\pt^{\ell\ell}\right)^2} + \sqrt{\mz^{2} + \left( \met \right)^2}}\: \right]^2 - \left\lvert \vecptll  + \vecmet \right\rvert^2 },
\label{eq::mt}
\end{equation*}
where \mz is the mass of the \Zboson boson, $\pt^{\ell\ell}$ is the transverse momentum of the
lepton pair and $\vecmet$ is the missing transverse momentum, with magnitude $\met$.
In the absence of such an excess, limits on the production rate of different
signal hypotheses are obtained from a simultaneous likelihood fit to the two mass distributions.
The first hypothesis is the ggF and VBF production of a heavy Higgs boson (spin-0 resonance) under
the narrow-width approximation (NWA).  The upper limits on the production rate of a heavy Higgs boson are then translated
into exclusion contours in the context of the two-Higgs-doublet model.  As several theoretical
models favour non-negligible natural widths, large-width assumption (LWA) models, assuming widths of 1\%, 5\%
and 10\% of the resonance mass, are also studied.  The interference between the heavy scalar and the
SM Higgs boson as well as between the heavy scalar and the
\ggZZ continuum background are taken into account in this study. Limits are also set on the Randall--Sundrum (RS) model \cite{rsModel1999, Davoudiasl200043} with a warped extra dimension giving rise to a spin-2 Kaluza--Klein (KK) excitation of the graviton \grav.

Other searches for diboson resonances decaying into  $WW$ or $ZZ$ or $WZ$ have been performed by ATLAS~\cite{Aaboud:2017itg,Aaboud:2017fgj,Aaboud:2017eta} and CMS~\cite{Sirunyan:2017nrt,Sirunyan:2017acf,Sirunyan:2017jtu}. 

With a significant increase in integrated luminosity and an improved discovery potential from the higher parton luminosities~\cite{Quigg:2009gg} at a centre-of-mass energy of $\sqrt{s}$ = 13~\tev\ as compared to $\sqrt{s}$ = 8~\tev, the results of this paper improve upon previous results published by the ATLAS Collaboration from a search for an additional heavy Higgs boson~\cite{HIGG-2013-20}. Results of a similar search from the data collected at the LHC with $\sqrt{s}$ = 8~\tev\ have also been reported by the CMS Collaboration~\cite{CMS-HIG-13-031}.

\section{ATLAS detector}
\label{sec:detector}
The ATLAS experiment is described in detail in Ref.~\cite{PERF-2007-01}. ATLAS is a multi-purpose detector with a forward–backward symmetric cylindrical geometry and a solid-angle\footnote{The ATLAS experiment uses a right-handed coordinate system with its origin at the nominal interaction point (IP) in the centre of the detector and the $z$-axis along the beam pipe. The $x$-axis points from the IP to the centre of the LHC ring, and the $y$-axis points upward. Cylindrical coordinates $(r,\phi)$ are used in the transverse plane, $\phi$ being the azimuthal angle around the $z$-axis. The pseudorapidity is defined in terms of the polar angle $\theta$ as $\eta=-\ln\tan(\theta/2)$.} coverage of nearly 4$\pi$.
The inner tracking detector (ID), covering the region $|\eta| <$ 2.5, consists of a silicon pixel detector, a silicon microstrip detector and a transition-radiation tracker. The innermost layer of the pixel detector, the insertable B-layer (IBL)~\cite{Capeans:1291633}, was installed between \RunOne and \RunTwo of the LHC. The inner detector is surrounded by a thin superconducting solenoid providing a $2\,\text{T}$ magnetic field, and by a finely segmented lead/liquid-argon (LAr) electromagnetic calorimeter covering the region $|\eta| <$ 3.2. A steel/scintillator-tile hadronic calorimeter provides coverage in the central region $|\eta| <$ 1.7. The end-cap and forward regions, covering the pseudorapidity range 1.5 $< |\eta| <$ 4.9, are instrumented with electromagnetic and hadronic LAr calorimeters, with steel, copper or tungsten as the absorber material.
A muon spectrometer (MS) system incorporating large superconducting toroidal air-core magnets surrounds the calorimeters. Three layers of precision wire chambers provide muon tracking in the range $|\eta| <$ 2.7, while dedicated fast chambers are used for triggering in the region $|\eta| <$ 2.4. The trigger system, composed of two stages, was upgraded~\cite{TRIG-PERF} before \RunTwo. The first stage, implemented with custom hardware, uses information from calorimeters and muon chambers to reduce the event rate from about 40 MHz to a maximum of 100~kHz. The second stage, called the high-level trigger (HLT), reduces the data acquisition rate to about 1~kHz on average. The HLT is  software-based and runs reconstruction algorithms similar to those used in the offline reconstruction.

\section{Data and Monte Carlo samples}
\label{sec:samples}
The proton--proton ($pp$) collision data used in these searches were collected 
by the ATLAS detector at a centre-of-mass energy of 13~\tev\  with a 25~ns 
bunch-spacing configuration during 2015 and 2016. The data are subjected to 
quality requirements: if any relevant detector component is not operating 
correctly during a period in which an event is recorded, the event is rejected. 
After these quality requirements, the total accumulated data sample corresponds 
to an integrated luminosity of \lumi.

Simulated events are used to determine the signal acceptance and some of the background contributions to these searches.
The particle-level events produced by each Monte Carlo (MC) event generator were processed through the ATLAS detector
simulation~\cite{SOFT-2010-01} within the $\textsc{Geant}~4$ framework~\cite{GEANT4}.  Additional inelastic $pp$ interactions (\pu) were overlaid on the simulated signal and background events. The MC event generator used for this is
\PYTHIAV{8.186}~\cite{Sjostrand:2007gs} with the A2 set of
tuned parameters~\cite{ATL-PHYS-PUB-2012-003} and the MSTW2008LO~\cite{Martin2009MSTW2008} parton distribution functions (PDF) set. The simulated events are weighted to reproduce the
observed distribution of the mean number of interactions per bunch crossing in data (\pu reweighting). 
The properties of the bottom and charm hadron decays were simulated by the \progname{EvtGen} v1.2.0
program~\cite{Lange:2001uf}. 

Heavy spin-0 resonance production was simulated using the \POWHEGBOX v2~\cite{Nason:2013ydw} MC event generator.  
Gluon--gluon fusion and vector-boson fusion production modes were calculated separately with matrix elements up to 
next-to-leading order (NLO) in QCD. \POWHEGBOX was interfaced to \PYTHIAV{8.212}~\cite{Sjostrand:2014zea} 
for parton showering and hadronisation, and for decaying the Higgs boson into the \htollll{} or \htollvv{} final states. 
The \progname{CT10} PDF set~\cite{Lai:2010vv} was used for the hard process.
Events from ggF and VBF production were generated in the 
$300~\gev < \mres < 1600$~\gev\ mass range under the NWA, using a step of 
100 (200)~\gev\ up to (above) 1000~\gev\ in mass. 
For the \llll final state, due to the sensitivity of the analysis at lower masses, events were also 
generated for $\mres = 200$~\gev. 

In addition, events from ggF production with a boson width of 5\%, 10\% and 15\% of the 
scalar mass \mres were generated with \MGMCatNLO v2.3.2~\cite{md52014} interfaced to \PYTHIAV{8.210} 
for parton showering and hadronisation for both final states. 
For the \llll final state, the \mfl distribution is parameterised analytically as described in 
Section~\ref{sec:h4l_sig_model}, and the samples with a width of 15\% of \mres are 
used to validate the parameterisation. For the \llvv final state, a reweighing procedure as described in Section~\ref{sec:h2l2v_sig_model} is used on fully simulated events to obtain the reconstructed \mt distribution at any value of mass and width tested.   
To have a better description of the jet multiplicity, 
\MGMCatNLO was also used to generate events for the process $pp \to H$ + $\geq$ 2 jets at NLO 
QCD accuracy with the FxFx merging scheme~\cite{Frederix2012}.

 The fraction of the ggF events that enter into the 
VBF-enriched category is estimated from the \MGMCatNLO simulation.

Spin-2 Kaluza--Klein gravitons from the Bulk Randall--Sundrum model~\cite{Agashe:2007zd} were generated with \MGMCatNLO at leading order (LO) in QCD. The dimensionless coupling $k/{\bar{M}_\mathrm{Pl}}$, where $\bar{M}_\mathrm{Pl} = M_\mathrm{Pl}/\sqrt{8\pi}$ 
is the reduced Planck scale and $k$ is the curvature scale of the extra dimension, is set to 1.
In this configuration, the width of the resonance is expected to be $\sim 6\%$ of its mass.

Mass points between 600~\gev\ and 2~\tev\ with 200~\gev\ spacing were generated for the \llvv final state. 
These samples were processed through a fast detector simulation~\cite{SOFT-2010-01} that uses a parameterisation of the response of electromagnetic 
and hadronic calorimeters~\cite{ATL-PHYS-PUB-2010-013}, while the response of the ID and MS detectors is fully simulated. 

The \qqZZ\ background for the \llvv final state was simulated by the \POWHEGBOX v2 event generator~\cite{Nason:2013ydw} and interfaced to \PYTHIAV{8.186}~\cite{Sjostrand:2007gs} for parton showering and hadronisation. 
The \progname{CT10nlo} PDF set~\cite{Lai:2010vv} was used for hard-scattering processes.
Next-to-next-to-leading-order (NNLO) QCD and NLO EW corrections are
included~\cite{Cascioli:2014yka,Grazzini:2015hta,Biedermann:2016yvs} as a
function of the invariant mass $m_{\itzz}$ of the \itzz\ system.  For the \llll final state, this background was simulated with
the \SHERPAV{v2.2.1}~\cite{Gleisberg:2008ta,Gleisberg:2008fv,Cascioli:2011va} event generator, with the
NNPDF3.0 NNLO PDF set~\cite{Ball:2014uwa} for the hard-scattering process.  NLO accuracy is achieved
in the matrix-element calculation for 0- and 1-jet final states and LO accuracy for 2- and 3-jet
final states. The merging with the \progname{Sherpa} parton
shower~\cite{Schumann:2007mg} was performed using the \progname{MePs}@NLO prescription~\cite{Hoeche:2012yf}.

NLO EW corrections were applied as a function of
$m_{\itzz}$~\cite{Biedermann:2016yvs,Biedermann:2016lvg}.  In addition, \SHERPAV{v2.2.1} was used for
the \llvv final state to scale the fraction of events in the VBF-enriched category obtained
from \POWHEGBOX simulation, because the \SHERPA event generator calculates matrix elements up to one
parton at NLO and up to three partons at LO.  
The EW production of a \itzz\ pair and two additional jets via vector-boson scattering
up to $O(\alpha_\mathrm{EW}^6)$ was generated using \SHERPA, where the process $ZZZ\to4\ell qq$
is also taken into account.

The \ggZZ\ production was modelled by \SHERPAV{v2.1.1} at LO in QCD for the \llll final state and by \progname{gg2VV}~\cite{Kauer:2015dma} for the \llvv final state, 
both including the off-shell \SMH boson contribution and the interference between the \SMH and \ZZ backgrounds. The K-factor accounting for higher-order QCD effects for the \ggZZ continuum production was
calculated for massless quark loops~\cite{Caola:2015psa,Campbell:2016ivq, Alioli:2016xab} in the
heavy-top-quark approximation~\cite{Melnikov:2015laa}, including the $gg\rightarrow
H^{*} \rightarrow ZZ$ process~\cite{Li:2015jva}. Based on these studies, a constant K-factor of 1.7
is used, and a relative uncertainty of 60\% is assigned to the normalisation in both searches.

The \WW and \WZ diboson events were simulated by \POWHEGBOX, using the \progname{CT10nlo} PDF set and \PYTHIAV{8.186} for parton showering and hadronisation. The production cross section of these samples is predicted at NLO in QCD.

Events containing a single $Z$ boson with associated jets were simulated using the \SHERPAV{v2.2.1} event generator. 
Matrix elements were calculated for up to two partons at NLO and four partons at LO using
the \progname{Comix}~\cite{Gleisberg:2008fv} and \progname{OpenLoops}~\cite{Cascioli:2011va} matrix-element 
generators and merged with the \SHERPA parton shower~\cite{Schumann:2007mg} using
the \progname{ME+PS@NLO} prescription~\cite{Hoeche:2012yf}.
The NNPDF3.0 NNLO PDF set was used in conjunction 
with dedicated parton-shower tuning developed by the \SHERPA authors. The \Zjets
events are normalised using the NNLO cross sections~\cite{GAVIN20112388}.

The triboson backgrounds $ZZZ$, $WZZ$, and $WWZ$ with fully leptonic decays and at least four prompt charged leptons were modelled 
using \SHERPAV{v2.1.1}. For the fully leptonic $t\bar{t}+Z$ background, with four prompt 
leptons originating from the decays of the top quarks and $Z$ boson, \MGMCatNLO was used.
The $t\bar{t}$ background, as well as the single-top and  $Wt$ production, were modelled using \POWHEGBOX v2 interfaced to \PYTHIAV{6.428}~\cite{pythia} with the Perugia 2012~\cite{Skands:2010ak} set of tuned parameters for parton showering and hadronisation, to \progname{PHOTOS}~\cite{Golonka:2005pn} for QED radiative corrections and to \progname{Tauola}~\cite{Jadach:1993hs,Golonka:2003xt} for the simulation of $\tau$-lepton decays. 

In order to study the interference treatment for the LWA case, samples containing the \ggZZ\ continuum background (\textit{B}) 
as well as its interference (\textit{I}) with a hypothetical heavy scalar (\textit{S}) were used 
and are referred to as \textit{SBI} samples hereafter.
In the \llll  final state the \progname{MCFM} NLO event generator~\cite{Campbell2011},  interfaced to \PYTHIAV{8.212}, was used to produce \textit{SBI} samples
where the width of the heavy scalar is set to 15\% of its mass, for masses of 200, 300, 400, 500, 600, 800, 1000, 1200 and 1400 \gev.
Background-only samples were also generated with the \progname{MCFM} event generator, and are used to extract the signal-plus-interference term (\textit{SI}) by subtracting them from the aforementioned \textit{SBI} samples.
For the \llvv final state, the  \textit{SBI} samples were generated with the \progname{gg2VV} event generator. 
The samples include signal events with a scalar mass of 400, 700, 900, 1200 and 1500 \gev.

\section{Event reconstruction}
\label{sec:objectreco}
Electrons are reconstructed using information from the ID and the electromagnetic calorimeter~\cite{PERF-2016-01}. 
Electron candidates are clusters of energy deposits associated with ID tracks, where the final track--cluster matching is performed after the tracks have been fitted with a Gaussian-sum filter (GSF) to account for bremsstrahlung energy losses. 
Background rejection relies on  the longitudinal and transverse shapes of the electromagnetic showers in the calorimeters, track--cluster matching and properties of tracks in the ID. 
All of this information, except for that related to track hits, is combined into a likelihood  discriminant. 

The selection used combines the likelihood with the number of track hits and defines two working points (WP) which are used in the analyses presented here.
The $\llll$ analysis uses a ``loose'' WP, with an efficiency ranging from 90\% for transverse momentum $\pt=20~\gev$ to 96\% for $\pt>60~\gev$. 
A ``medium'' WP was chosen for the $\llvv$ analysis with an efficiency increasing from 82\% at $\pt=20~\gev$ to 93\% for $\pt>60~\gev$. The electron's transverse momentum is computed from the cluster energy and the track direction at the interaction point. 
  
Muons are formed from tracks reconstructed in the ID and MS, 
and their identification is primarily based on the presence of the track or track segment in the MS~\cite{PERF-2015-10}.  
If a complete track is present in both the ID and the MS, a combined muon track is formed by a global fit using the hit information from both the ID 
and MS detectors (combined muon), otherwise the momentum is measured using the ID, and the MS track segment serves as identification (segment-tagged muon). 
The segment-tagged muon is limited to the centre of the barrel region ($|\eta| <0.1$) which has reduced MS geometrical coverage. 
Furthermore, in this central region an ID track with \pt $ > 15$ \gev\ is identified as a muon if its calorimetric energy deposition is consistent with a minimum-ionising particle (calorimeter-tagged muon). 
In the forward region ($2.5 < |\eta| < 2.7$) with limited or no ID coverage, 
the MS track is either used alone (stand-alone muon) or combined with silicon hits, if found in the forward ID (combined muon).
The ID tracks associated with the muons are required to have a minimum number of associated hits in each of the ID subdetectors to ensure good track reconstruction. 
The stand-alone muon candidates are required to have hits in each of the three MS stations they traverse.
A ``loose'' muon identification WP, which uses all muon types and has an efficiency of 98.5\%, is adopted by the $\llll$ analysis. 
For the $\llvv$ analysis a ``medium'' WP is used, which only includes combined muons and has an efficiency of 97\%.

Jets are reconstructed using the anti-$k_t$ algorithm~\cite{Cacciari:2008gp} with a radius parameter $R$ = 0.4 implemented in the \progname{FastJet} package~\cite{Cacciari:2011ma}, and positive-energy clusters of calorimeter cells as input. 
The algorithm suppresses noise and \pu by keeping only cells with a significant energy deposit and their neighbouring cells. 
Jets are calibrated using a dedicated scheme designed to adjust, on average, the energy measured in the calorimeter to that of the true jet energy~\cite{PERF-2016-04}. 
The jets used in this analysis are required to satisfy $\pt > 20~\GeV$ and $|\eta |<4.5$. 
To reduce the number of jet candidates originating from \pu vertices, an additional requirement that uses the track and vertex information inside a jet
is imposed on jets with $\pt < 60~\GeV$ and $|\eta |<2.4$~\cite{PERF-2014-03}.

Jets containing $b$-hadrons, referred to as $b$-jets, are identified by the long lifetime,
high mass and decay multiplicity of $b$-hadrons, as well as the hard $b$-quark fragmentation function. 
The $\llvv$ analysis identifies $b$-jets of $\pt > 20~\GeV$ and $|\eta |<2.5$ using an algorithm that achieves an identification 
efficiency of about 85\% in simulated $t\bar{t}$ events, 
with a rejection factor for light-flavour jets of about 33~\cite{PERF-2012-04, ATL-PHYS-PUB-2016-012}.

Selected events are required to have at least one vertex with two associated tracks with \pt\ $>$ 400 \mev, 
and the primary vertex is chosen to be the vertex reconstructed  with the largest $\sum \pt^2$. 
As lepton and jet candidates can be reconstructed from the same detector information, 
a procedure to resolve overlap ambiguities is applied. 
If an electron and a muon share the same ID track, 
the muon is selected unless it is calorimeter-tagged and does not have a MS track, 
or is a segment-tagged muon, in which case the electron is selected.  
Reconstructed jets which overlap with electrons (muons) in a cone of size $\Delta R\equiv \sqrt{(\Delta \eta)^2 + (\Delta \phi)^2}= $ 0.2 (0.1) are removed. 

The missing transverse momentum \vecmet, which accounts for the imbalance of visible momenta in the plane transverse to the beam axis, is computed as the negative vector sum of the transverse momenta of all identified electrons, muons and jets, as well as a ``soft term'', accounting for unclassified soft tracks and energy clusters in the calorimeters~\cite{PERF-2014-04}. 
This analysis uses a track-based soft term, which is built by combining the information provided by the ID and the calorimeter, in order to minimise the effect of \pu which degrades the \met resolution. The soft term is computed using the momenta of the tracks associated with the primary vertex, while the jet and electron momenta are computed at the calorimeter level to allow the inclusion of neutral particles.
Jet--muon overlap is accounted for in the \met calculation. 
This corrects for fake jets due to \pu close to muons and double-counted jets from muon energy losses. 

\section{\htollll\ event selection and background estimation}
\label{sec:h4lanal}
\subsection{Event selection}
\label{sec:sel4l}
Four-lepton events are selected and initially classified according to the lepton flavours:
$4\mu$, $2e2\mu$, $4e$, called ``channels'' hereafter. 
 They  are selected with single-lepton, dilepton and trilepton triggers, 
with the dilepton and trilepton ones including electron(s)--muon(s) triggers. 
Single-electron triggers apply ``medium'' or ``tight''  likelihood identification, 
whereas multi-electron triggers apply ``loose'' or ``medium'' identification.  
For the bulk of the data, recorded in 2016, the lowest \pt\ threshold for the single-electron 
(muon) triggers used is set to 26 (26)~\gev, for the dielectron (dimuon) triggers to 15 (10)~\gev\ 
and for the trielectron (trimuon) triggers to 12 (6)~\gev. For the data collected in 2015, 
the instantaneous luminosity was lower so the trigger thresholds were lower;
this increases the signal efficiency by less than 1\%. 
Globally, the trigger efficiency for signal events passing the final selection requirements is about 98\%.  

In each channel, four-lepton candidates are formed by selecting 
a lepton-quadruplet made out of two same-flavour, opposite-sign lepton pairs, selected as 
described in Section~\ref{sec:objectreco}. Each electron (muon) must satisfy 
$\pt > 7$ (5) \gev\ and be measured in the pseudorapidity range of 
$\left|\eta \right|<2.47$ ($2.7$). The highest-\pt\ lepton in the quadruplet 
must satisfy \pt $> 20$~\gev, and the second (third) lepton in \pt\ order must satisfy 
\pt $> 15$~\gev\ ($10$~\gev). In the case of muons, at most one calorimeter-tagged, 
segment-tagged or stand-alone ($2.5 < |\eta| < 2.7$) muon is allowed per quadruplet.

If there is ambiguity in assigning leptons to a pair, only one quadruplet per channel is selected by keeping the quadruplet with the 
lepton pairs closest (leading pair) and second closest (subleading pair) to 
the $Z$ boson mass, with invariant masses referred to as  $m_{12}$ and $m_{34}$ respectively. In the selected quadruplet, $m_{12}$ is required to be $50~\gev < m_{12} < 106~\gev$, 
while $m_{34}$ is required to be less than 115~\gev\ and greater than a threshold
that is 12~\gev\ for $\mfl \leq 140~\gev$, rises linearly from 12~\gev\ to 50~\gev\ with \mfl in 
the interval of [140~\gev, 190~\gev] and is fixed to 50~\gev\ for $\mfl > 190~\gev$.

Selected quadruplets are required to have their leptons separated from each other by $\Delta R>0.1$ if they are of the same flavour and by $\Delta R>0.2$ otherwise.  
For $4\mu$ and $4e$ quadruplets, if an opposite-charge same-flavour lepton pair is found 
with $m_{\ell\ell}$ below 5 \gev, the quadruplet is removed to suppress the contamination from $J/\psi$ mesons. If multiple quadruplets from different channels are selected at this point, 
only the quadruplet from the channel with the highest expected signal rate is retained, in the order: $4\mu$, $2e2\mu$, $4e$. 

The \Zjets and \ttbar background contributions are reduced by imposing impact-parameter requirements
as well as track- and calorimeter-based isolation requirements on the leptons. The transverse impact-parameter 
significance, defined as the impact parameter calculated with respect to the measured beam
line position in the transverse plane divided by its uncertainty, $|d_0|/\sigma_{d_0}$, for all
muons (electrons) is required to be lower than 3 (5). The normalised track-isolation discriminant,
defined as the sum of the transverse momenta of tracks, inside a cone of size $\Delta R=0.3\ (0.2)$
around the muon (electron) candidate, excluding the lepton track, divided by the lepton \pt, is
required to be smaller than 0.15. The larger muon cone size corresponds to that used by the muon
trigger.  Contributions from \pu are suppressed by requiring tracks in the cone to originate
from the primary vertex. To retain efficiency at higher \pt, the track-isolation cone size is
reduced to 10~\gev/\pt for \pt above 33 (50)~\gev\ for muons (electrons).

The relative calorimetric isolation is computed as the sum of the cluster transverse energies $\et$,
in the electromagnetic and hadronic calorimeters, with a reconstructed barycentre inside a cone of
size $\Delta R=0.2$ around the candidate lepton, divided by the lepton $\pt$.  The clusters used for
the isolation are the same as those for reconstructing jets.  The relative calorimetric isolation is
required to be smaller than 0.3 (0.2) for muons (electrons). The measured calorimeter energy around
the muon (inside a cone of size $\Delta R=0.1$) and 
the cells within $0.125 \times 0.175$ in $\eta \times \phi$ around the electron 
barycentre are excluded from the respective sums.  The \pu and underlying-event contributions to
the calorimeter isolation are subtracted event by event~\cite{Cacciari:2007fd}.  For both the track-
and calorimeter-based isolation requirements, any contribution arising from other leptons of the
quadruplet is subtracted.

An additional requirement based on a vertex-reconstruction algorithm, which fits  the four-lepton 
candidates with the constraint that they originate from a common vertex, is applied
in order to further reduce the  $Z+\mathrm{jets}$ and $t\bar{t}$ background contributions. 
A loose cut of $\chi^{2} / \mathrm{ndof} < 6$ for $4\mu$ and $<9$ for the other channels is applied,
which retains a signal efficiency larger than  $99 \%$ in all channels.

The QED process of radiative photon production in $Z$ boson decays is well modelled by simulation. 
Some of the final-state-radiation (FSR) photons can be identified in the calorimeter and incorporated into 
the \llll\ analysis.  The strategy to include FSR photons into the reconstruction of $Z$ bosons is 
the same as in \runa~\cite{HIGG-2013-20}. It consists of a search for collinear (for muons) 
and non-collinear FSR photons (for muons and electrons) with only one FSR photon allowed per event. 
After the FSR correction, the lepton four-momenta of both dilepton pairs are recomputed by 
means of a $Z$-mass-constrained kinematic fit. The fit uses a Breit--Wigner $Z$ boson line-shape and 
a single Gaussian function  per lepton to model the momentum response function with the Gaussian width 
set to the expected resolution for each lepton. The $Z$-mass constraint is applied to both $Z$ 
candidates, and improves the $m_{4\ell}$ resolution by about 15\%. 

In order to be sensitive to the VBF production mode, events are classified into four categories:
one for the VBF production mode and three for the ggF production mode, one for each of the three channels.
If an event has two or more jets with \pt greater than $30$~\gev, 
with the two leading  jets being well separated in $\eta$, $\detajj > 3.3$, and having an
invariant mass $\mjj > 400~\gev$, this event is classified into the VBF-enriched category; 
otherwise the event is classified into one of the ggF-enriched categories.
Such classification is used only in the search for a heavy scalar produced with the NWA.

The signal acceptance, defined as the ratio of the number of reconstructed events passing the
analysis requirements to the number of simulated events in each category, is shown in
Table~\ref{tab:H4lacc}, for the ggF and VBF production modes as well as different resonance masses.
The contribution from final states with $\tau$ leptons decaying into electrons or muons is found to be
negligible.
\begin{table}[htbp!]
  \caption{\label{tab:H4lacc}Signal acceptance for the \llll\ analysis, for both the ggF and VBF
    production modes and resonance masses of 300 and 600~\gev. The acceptance is defined as the
    ratio of the number of reconstructed events after all selection requirements to the number of
    simulated events for each channel/category.  }
  \tiny \centering \resizebox{\textwidth}{!}{
    \begin{tabular}{ c c r r r r }
      \toprule
      \multirow{2}{*}{Mass} & \multirow{2}{*}{Production mode} & \multicolumn{3}{c}{ggF-enriched categories}  & \multirow{2}{*}{VBF-enriched category} \\
      & & $4\mu$ channel & $2e2\mu$ channel & $4e$ channel & \\
      \midrule
      \multirow{2}{*}{300~\gev} & ggF & 56\% & 48\% & 40\% & 1\% \\ 
      & VBF & 36\% & 30\% & 24\% & 21\% \\ 
      \midrule
      \multirow{2}{*}{600~\gev} & ggF & 64\% & 56\% & 48\% & 3\% \\ 
      & VBF & 36\% & 34\% & 32\% & 26\% \\ 
      \bottomrule
    \end{tabular}
  }
\end{table}

\subsection{Background estimation}
\label{sec:bkg4l}
The main background component in the \htollll\ final state, accounting for 97\% of the total
expected background events, is non-resonant \itzz\ production. This arises from
quark--antiquark annihilation (86\%), gluon-initiated production (10\%) and a small contribution from EW vector-boson scattering (1\%).
The last is more important in the VBF-enriched category, where it accounts for 16\% of the total expected background. 
These backgrounds are all modelled by MC simulation as
described in Section~\ref{sec:samples}.  Additional background comes from the \Zjets and \ttbar
processes, which contribute at the percent level and decrease more rapidly than the non-resonant
\itzz\ production as a function of \mfl. 
These backgrounds are estimated using data where possible, following slightly different approaches for
final states with a dimuon (\llmumu) or a dielectron (\llee) subleading pair~\cite{HIGG-2013-21}.

The \llmumu\ non-\ZZ background comprises mostly \ttbar and \Zjets events, where in the latter
case the muons arise mostly from heavy-flavour semileptonic decays and to a lesser extent from
$\pi$/$K$ in-flight decays. The contribution from single-top production is negligible. The
normalisations of the \Zjets\ and \ttbar\ backgrounds are determined using fits to the invariant
mass of the leading lepton pair in dedicated data control regions.  The control regions are formed
by relaxing the $\chi^2$ requirement on the vertex fit, and by inverting and relaxing isolation
and/or impact-parameter requirements on the subleading muon pair. An additional control region
($e\mu \mu\mu$) is used to improve the \ttbar\ background estimate. Transfer factors to extrapolate
from the control regions to the signal region are obtained separately for \ttbar\ and \Zjets\ using
simulated events. The transfer factors have a negligible impact on the \mfl shape of the \llmumu\ background. 

The main background for the \llee\ process arises from the misidentification of light-flavour jets as
electrons, photon conversions and the semileptonic decays of heavy-flavour hadrons.  The
\llee\ control-region selection requires the electrons in the subleading lepton pair to have the
same charge, and relaxes the identification and isolation requirements on the electron candidate,
denoted $X$, with the lower transverse momentum.  The heavy-flavour background is completely
determined from simulation, whereas the light-flavour and photon-conversion background is obtained
with the sPlot~\cite{Pivk2005356} method, based on a fit to the number of hits in the innermost ID
layer in the data control region.  
Transfer factors for the light-flavour jets and
converted photons, obtained from simulated samples, are corrected using a $Z+X$ control region and
then used to extrapolate the extracted yields to the signal region.  Both the yield extraction and
the extrapolation are performed in bins of the transverse momentum of the electron candidate and the
jet multiplicity.

The $WZ$ production process is included in the data-driven estimates for the \llee\ final states, while it
is added from simulation for the \llmumu\ final states. The contributions from \ttv\ (where $V$ stands for either a $W$ or a $Z$ boson) and triboson
processes are minor and taken from simulated samples.

\subsection{Signal and background modelling}
\label{sec:h4l_sig_model}
The parameterisation of the reconstructed four-lepton invariant mass \mfl distribution 
for signal and background is based on the MC simulation and used to fit the data. 

\begin{figure}[tbp]
  \subfigure[\label{fig:param_sub1}]{\includegraphics[width=0.49\textwidth]{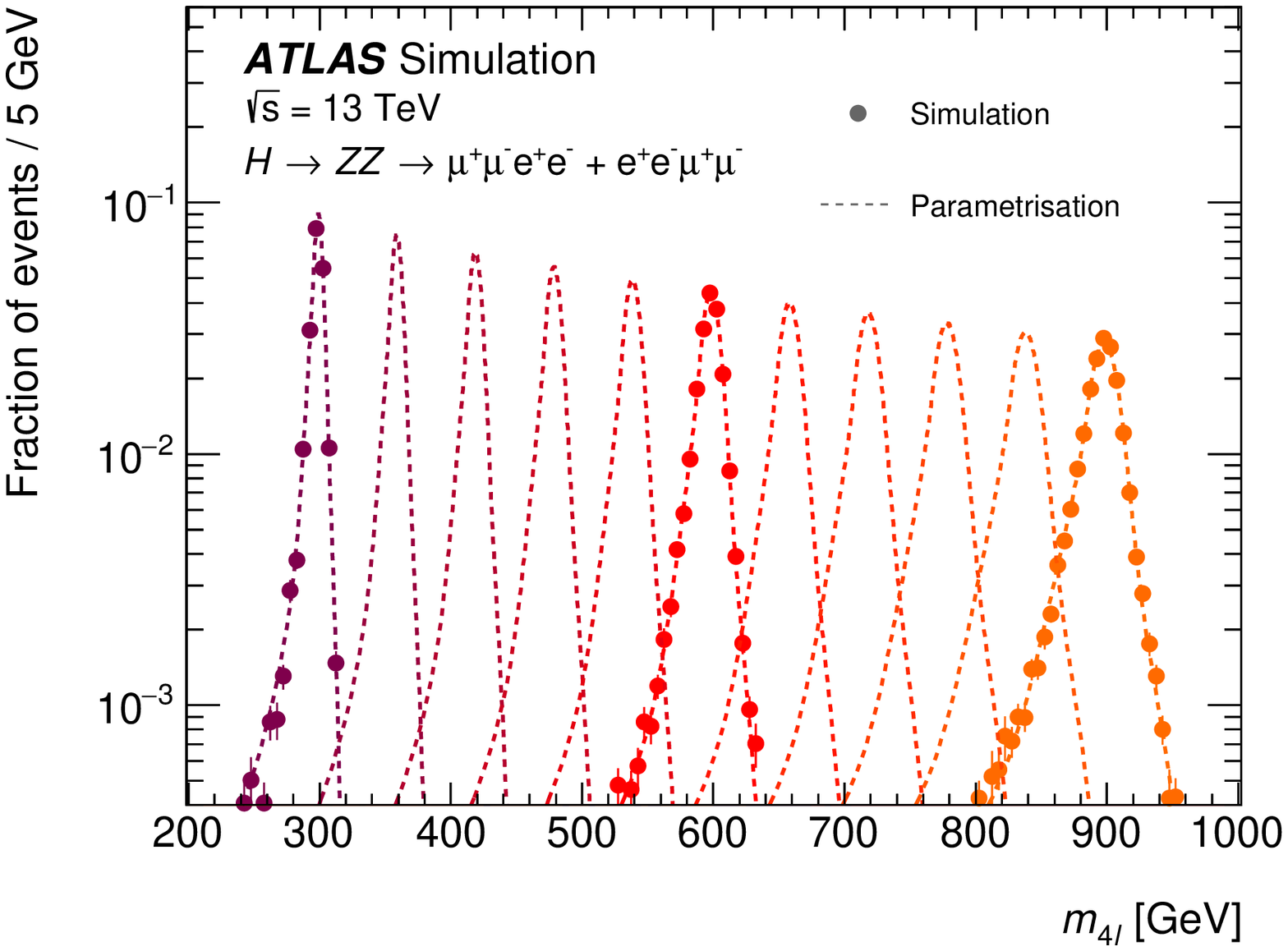}}
  \subfigure[\label{fig:param_sub2}]{\includegraphics[width=0.49\textwidth]{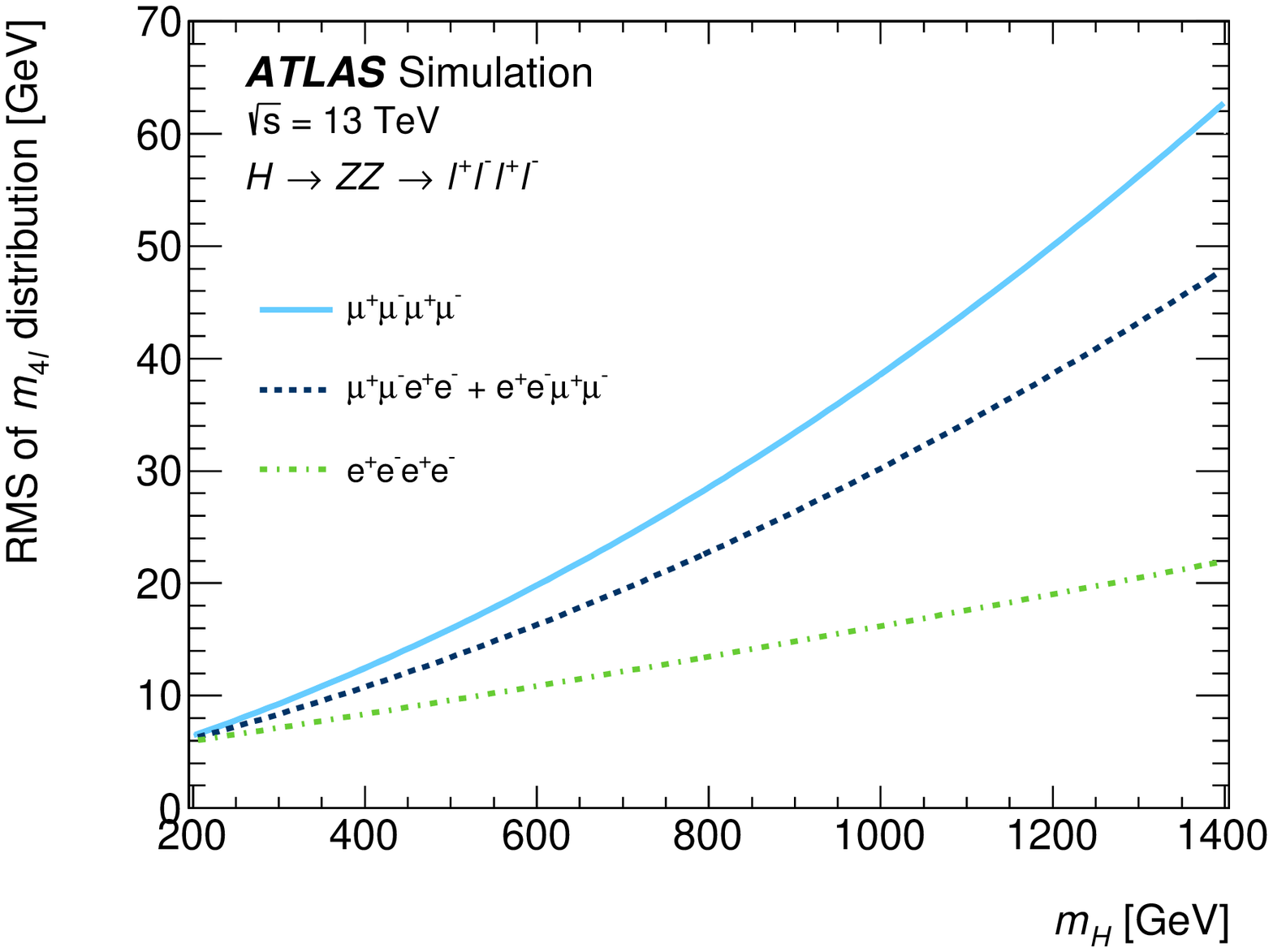}}
  \caption{ \subref{fig:param_sub1} Parameterisation of the four-lepton invariant mass (\mfl) spectrum for various resonance mass (\mres) hypotheses in the NWA. Markers show the simulated \mfl\ distribution for three specific values of \mres\ (300, 600, 900~\gev), normalised to unit area, and the dashed lines show the parameterisation used in the   $2e2\mu$ channel
  for these mass points as well as for intervening ones. \subref{fig:param_sub2} RMS of the four-lepton invariant mass distribution  as a function of \mres.} 
  \label{fig:param}
\end{figure}

In the case of a narrow resonance, the width in \mfl is determined 
by the detector resolution, which is modelled by the sum of a Crystal Ball 
($\mathcal{C}$) function~\cite{CrystalBall1,CrystalBall2} and a Gaussian ($\mathcal{G}$) function:  
$$ P_{s} (\mfl) = f_{\mathcal{C}} \times \mathcal{C}(\mfl; \mu, \sigma_{\mathcal{C}}, \alpha_{\mathcal{C}}, n_{\mathcal{C}}) + (1 - f_{\mathcal{C}}) \times \mathcal{G}(\mfl; \mu, \sigma_{\mathcal{G}}). $$
The Crystal Ball and the Gaussian functions share the same peak value of \mfl ($\mu$),
but have different resolution parameters, $\sigma_{\mathcal{C}}$ and $\sigma_{\mathcal{G}}$.
The $\alpha_{\mathcal{C}}$ and $n_{\mathcal{C}}$ parameters control the shape 
and position of the non-Gaussian tail and the parameter $f_{\mathcal{C}}$ 
ensures the relative normalisation of the two probability density functions. 
To improve the stability of the parameterisation in the full mass range considered, the parameter $n_{\mathcal{C}}$ is set to a fixed value. 
The bias in the extraction of signal yields introduced by using the analytical function is below 1.5\%. 
The function parameters are determined separately for each final state using signal simulation,
and fitted to first- and second-degree polynomials in scalar mass \mres 
to interpolate between the generated mass points. 
The use of this parameterisation for the function parameters introduces an extra bias 
in the signal yield and \mres extraction of about 1\%.  
An example of this parameterisation is illustrated in Figure~\ref{fig:param}, 
where the left plot shows the mass distribution for simulated samples at \mres $=300,600,900$~\gev\ 
and the right plot shows the RMS of the \mfl distribution in the range considered for this search.

In the case of the LWA, the particle-level line-shape of \mfl is derived 
from a theoretical calculation, as described in Ref.~\cite{Goria:2011wa}, 
and is then  convolved with the detector resolution, using the same 
procedure as for the modelling of the narrow resonance.

The \mfl distribution for the \ZZ continuum background is taken from MC simulation, 
and parameterised by an empirical function for both the quark- and gluon-initiated processes: 
\begin{equation*}
f_{qqZZ/ggZZ}(m_{4\ell}) = (f_{1} (m_{4\ell})+ f_{2} (m_{4\ell})) \times H(m_{0}-m_{4\ell}) \times C_0 + f_{3} (m_{4\ell}) \times H(m_{4\ell}-m_{0}),
\label{eq::bkg_model}
\end{equation*}
where:
\begin{equation*}
  \begin{split}
    f_{1} (m_{4\ell}) &= \exp(a_1 + a_2 \cdot m_{4\ell}),  \\
        f_{2} (m_{4\ell}) &= \left\{ \frac{1}{2} + \frac{1}{2}\,\text{erf}\left(\frac{m_{4\ell}-b_1}{b_2}\right)\right\} \times \frac{1}{1+\exp\left(\frac{m_{4\ell}-b_1}{b_3}\right)}, \\
    f_{3} (m_{4\ell}) &=  \exp\left(c_1 + c_2 \cdot m_{4\ell} + c_3 \cdot m_{4\ell}^2  + c_4 \cdot m_{4\ell}^{2.7}\right), \\
    C_0 &= \frac{f_{3} (m_0)} {f_{1} (m_0)+f_{2} (m_0)}.
  \end{split}
  \label{eq::bkg_model2}
\end{equation*}

The function's first part, $f_1$, covers the low-mass part of the spectrum where 
one of the $Z$ bosons is off-shell, while $f_2$ models the \ZZ threshold around 2$\cdot m_Z$ 
and $f_3$ describes the high-mass tail. 
The transition between low- and high-mass parts is performed by the Heaviside 
step function $H(x)$ around $m_0 = 240$~\gev. The continuity of the function 
around $m_0$ is ensured by the normalisation factor $C_0$ that is applied 
to the low-mass part. Finally, $a_i$, $b_i$ and $c_i$ are shape parameters which 
are obtained by fitting the \mfl distribution in simulation for each category.
The uncertainties in the values of these parameters from the fit are found to be negligible. 
The MC statistical uncertainties in the high-mass tail are taken into account by 
assigning a 1\% uncertainty to $c_4$.

The \mfl shapes  are extracted from simulation for most background components (\ttv, \vvv, \llmumu\ and  heavy-flavour hadron component of \llee),  except for the light-flavour  jets and photon conversions in the case  of \llee\ background,  
which is taken from the control region as described in Section~\ref{sec:bkg4l}.

\subsubsection*{Interference modelling}
\label{sec:h4l_interf_model}
The gluon-initiated production of a heavy scalar \BigH, the SM \SMH and the \ggZZ continuum background
all share the same initial and final state, and thus lead to interference terms in
the total amplitude.  Theoretical calculations described in~Ref.~\cite{Sig-bkg_inter} have shown that the
effect of interference could modify the integrated cross section by up to $\mathcal{O}$(10\%),
and this effect is enhanced as the width of the heavy scalar increases. Therefore, a search for a heavy
scalar Higgs boson in the LWA case must properly account for two interference effects: the
interference between the heavy scalar and the SM Higgs boson (denoted by \IntHh) and between the
heavy scalar and the \ggZZ continuum (denoted by \IntHB).

Assuming that \BigH and \SMH bosons have similar properties, as postulated by the 2HDM, they have the same  production and decay amplitudes and therefore the only difference between the signal and interference terms in the production cross section  comes from the propagator. Hence, the acceptance and resolution of the signal and interference terms are expected to be the same. The \IntHh interference is obtained by reweighting the particle-level line-shape of generated signal events using the following formula:

\begin{equation*}
w(\mfl) = \frac{2 \cdot Re \left [\frac{1}{s-s_H} \cdot  \frac{1}{(s-s_h)^*} \right ] }{\frac{1}{\left | s - s_H \right |^2}},
\end{equation*}
where $1/\left(s-s_{H (h)}\right)$ is the propagator for a scalar (\BigH or \SMH). The particle-level line-shape is then convolved  with the detector resolution function, and the signal and interference acceptances are assumed to be the same.

In order to extract the \IntHB interference contribution, signal-only and background-only samples are subtracted from the generated \textit{SBI} samples. The extracted particle-level \mfl distribution for the \IntHB interference term is then convolved with the detector resolution.

\begin{figure}[tbp]
\begin{center}
\includegraphics[width=.9\textwidth]{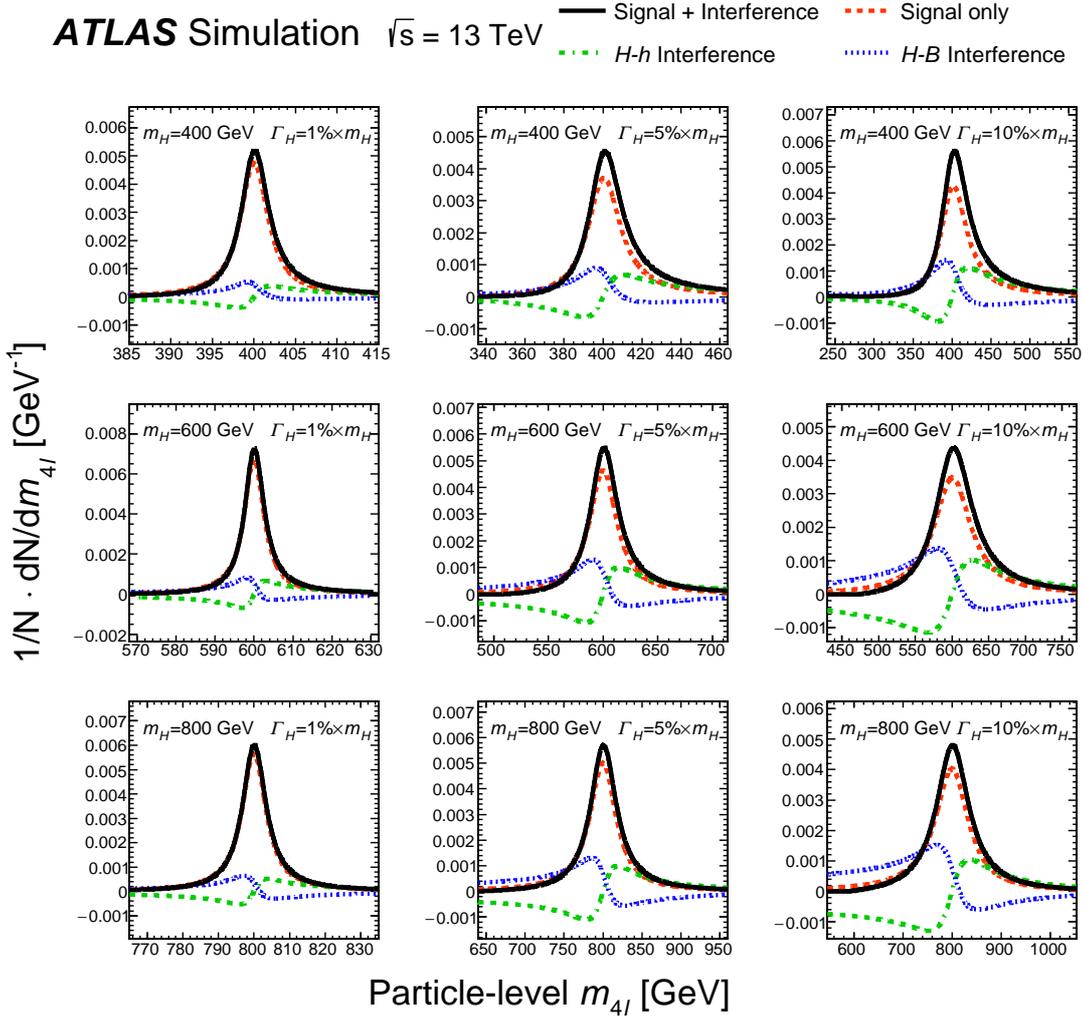}
\end{center}
\caption{Particle-level four-lepton mass \mfl\ model for signal only (red), \IntHh interference
  (green), \IntHB interference (blue) and the sum of the three processes (black). Three values of
  the resonance mass \mres (400, 600, 800~\gev) are chosen, as well as three values of the resonance
  width $\Gamma_{H}$ ($1\%$, $5\%$, $10\%$ of \mres). The signal cross section, which determines the
  relative contribution of the signal and interference, is taken to be the cross section of the
  expected limit for each combination of \mres\ and $\Gamma_{H}$. The full model (black) is finally
  normalised to unity and the other contributions are scaled accordingly.\label{fig:LWA_Total_SM} }
\end{figure}

Figure~\ref{fig:LWA_Total_SM} shows the overlay of the signal, both interference effects and the
total line-shape for different mass and width hypotheses assuming the couplings expected in the SM for a
heavy Higgs boson.   As can be seen, the two interference effects tend to cancel out, and the total interference yield is for the most part positive, 
enhancing the signal.

\label{sec:model4l}

\section{\htollvv\ event selection and background estimation}
\label{sec:hllvvanal}
\subsection{Event selection}
\label{sec:selllvv}
The analysis is designed to select \zztollvv events (with $\ell = e,\mu$), where the missing neutrinos are identified by a large \met, 
and to discriminate against the large \Zjets, \WZ and top-quark backgrounds.

Events are required to pass either a single-electron or a single-muon trigger, where
 different \pt thresholds are used depending on the instantaneous luminosity of the LHC\@. 
For the 2015 data the electron and muon triggers had \pt thresholds of 24~\gev\ and 20~\gev\ respectively,
while for 2016 the muon trigger threshold was increased to 24 \GeV.
For both triggers, the threshold is set to 26 \GeV\ when the instantaneous luminosity exceeds the value of $10^{34}$ \instlumi.
The trigger efficiency for signal events passing the final selection is about 99\%.

Events are selected if they contain exactly two opposite-charge leptons of the same flavour
and ``medium'' identification, with the more energetic lepton having $\pt > 30$~\gev\ and the other one having $\pt > 20$~\gev.
The same impact-parameter significance criteria as defined in Section~\ref{sec:sel4l} are applied to the selected leptons. 
Track- and calorimeter-based isolation criteria as defined in Section~\ref{sec:sel4l} are also applied to the leptons, but in this analysis the isolation criteria are optimised by adjusting the isolation threshold so that their selection efficiency is 99\%. 
If an additional lepton with $\pt > 7$ \gev\ and ``loose'' identification is found, 
the event is rejected to reduce the amount of \WZ background.
In order to select leptons originating from the decay of a \Zboson boson, 
the invariant mass of the pair is required to be in the range 76 to 106 \GeV.
Moreover, since a \Zboson boson originating from the decay of a high-mass particle is boosted, 
the two leptons are required to be produced with an angular separation of $\Delta R_{\ell\ell}<1.8$.

Events with neutrinos in the final state are selected by requiring $\met > 120$ \GeV, 
and this requirement heavily reduces the amount of \Zjets background. 
In signal events with no initial- or final-state radiation the visible \Zboson boson's transverse momentum 
is expected to be opposite the missing transverse momentum,
and this characteristic is used to further suppress the \Zjets background.
The azimuthal angle between the dilepton system and the missing transverse 
momentum ($\Delta\phi(\ell\ell,\vecmet)$)  is thus required to be greater than 2.7 
and the fractional \pt difference, defined as \fracpt, to be less than 20\%, where \ptmissjet.

Additional selection criteria are applied to keep only events with \met originating from neutrinos rather than 
detector inefficiencies, poorly reconstructed high-\pt muons or mismeasurements in the hadronic calorimeter.
If at least one reconstructed jet has a \pt greater than 100 \GeV, 
the azimuthal angle between the highest-\pt jet and the missing transverse momentum
is required to be greater than 0.4. Similarly, if \met is found to be less than 40\% 
of the scalar sum of the transverse momenta of leptons and jets in the event (\HT), the event is rejected.
Finally, to reduce the \ttbar background, events are rejected whenever a $b$-tagged jet is found.  

The sensitivity of the analysis to the VBF production mode is increased by creating a dedicated category of VBF-enriched events. 
The selection criteria, determined by optimising the expected signal significance using signal and background MC samples, require the presence of at least two jets with $\pt > 30$~\gev\, where the two highest-$\pt$ jets are widely separated in $\eta$, $\detajj > 4.4$, and have an invariant mass \mjj greater than $550$~\gev.

The signal acceptance, defined as the ratio of the number of reconstructed events passing the
analysis requirements to the number of simulated events in each category, is
shown in Table~\ref{tab:H2l2vacc}, for the ggF and VBF production modes as well as for different
resonance masses. The acceptance increases with mass due to a kinematic threshold determined by the \met selection criteria. Hence the \llvv\ search considers only masses of 300~\gev\ and above, where its inclusion improves the combined sensitivity.  \begin{table}[t]
  \caption{\label{tab:H2l2vacc}Signal acceptance for the \llvv\ analysis, for both the ggF and VBF
    production modes and resonance masses of 300 and 600~\gev. The acceptance is defined as the
    ratio of the number of reconstructed events after all selection requirements to the number of
    simulated events for each channel/category.  }
  \tiny \centering \resizebox{\textwidth}{!}{
    \begin{tabular}{ c c r r r }
      \toprule
      \multirow{2}{*}{Mass} & \multirow{2}{*}{Production mode} & \multicolumn{2}{c}{ggF-enriched categories}  & \multirow{2}{*}{VBF-enriched category} \\
      & & \mm channel & \ee channel & \\
      \midrule
      \multirow{2}{*}{300~\gev} & ggF & 6\% & 5\% & <0.05\% \\ 
      & VBF & 2.6\% & 2.4\% & 0.7\% \\ 
      \midrule
      \multirow{2}{*}{600~\gev} & ggF & 44\% & 44\% & 1\% \\ 
      & VBF & 27\% & 27\% & 13\% \\ 
      \bottomrule
    \end{tabular}
  }
\end{table}

\subsection{Background estimation}
\label{sec:sigbkgllvv}
The dominant and irreducible background for this search is non-resonant \ZZ production, which accounts for about 60\% of the expected background events.
The second largest background comes from \WZ production ($\sim$30\%) followed by \Zjets production with poorly reconstructed \met ($\sim$6\%). 
Other sources of background are the \WW, \ttbar, \Wt and \Ztt processes ($\sim$3\%). 
Finally, a small contribution comes from \Wjets, \ttbar, single-top-quark and multi-jet processes, 
with at least one jet misidentified as an electron or muon, as well as from \ttv/\vvv events. In both the ggF- and in the VBF-enriched signal regions, the \ZZ background is modelled   
 using MC simulation and normalised using SM predictions, as explained in Section~\ref{sec:samples}. 
The remaining backgrounds are mostly estimated using control samples in data.

\begin{figure}[tbp]
  \subfigure[\label{fig:llvv_crs_sub1}]{\includegraphics[width=0.49\textwidth]{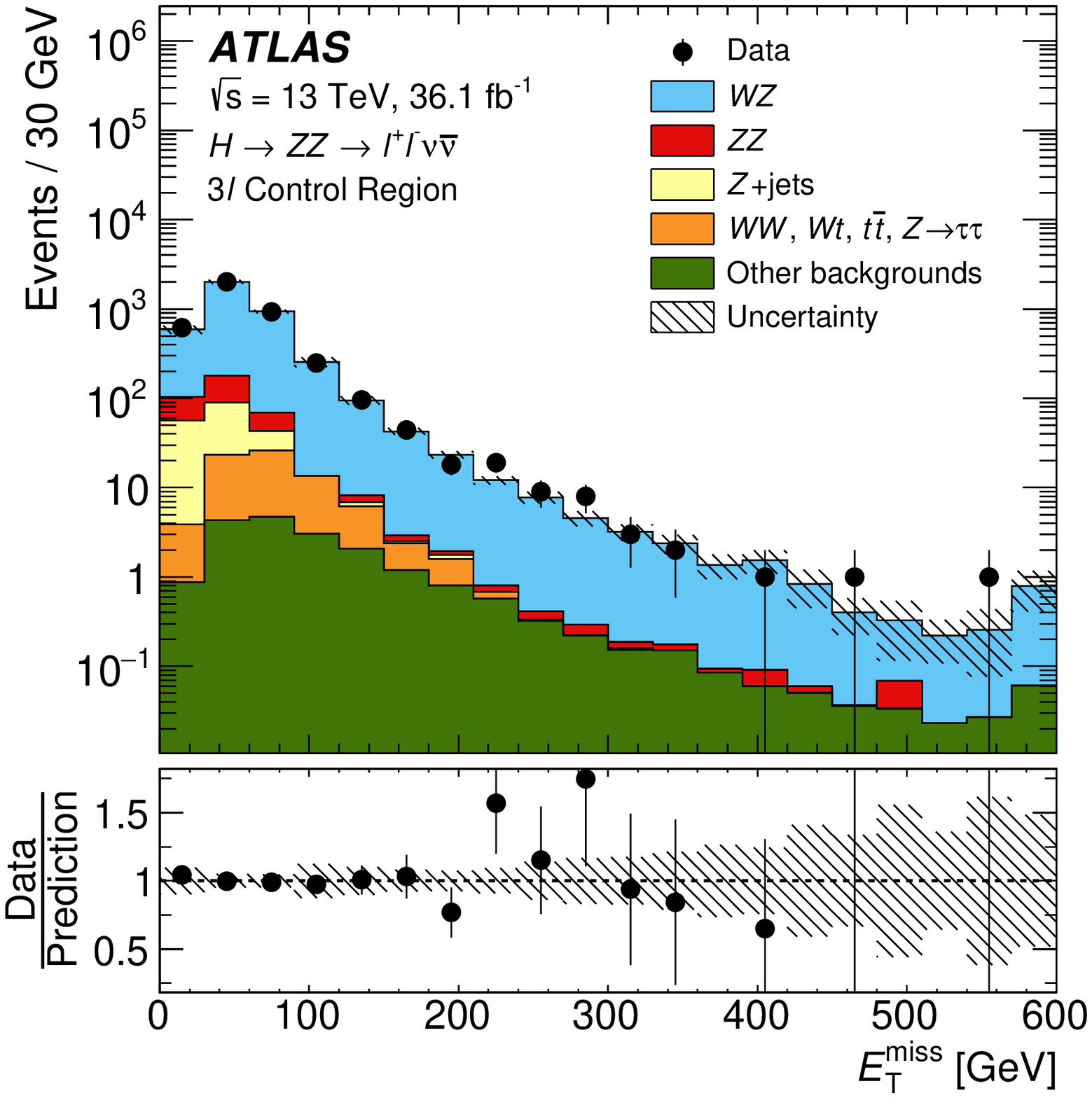}}
  \subfigure[\label{fig:llvv_crs_sub2}]{\includegraphics[width=0.49\textwidth]{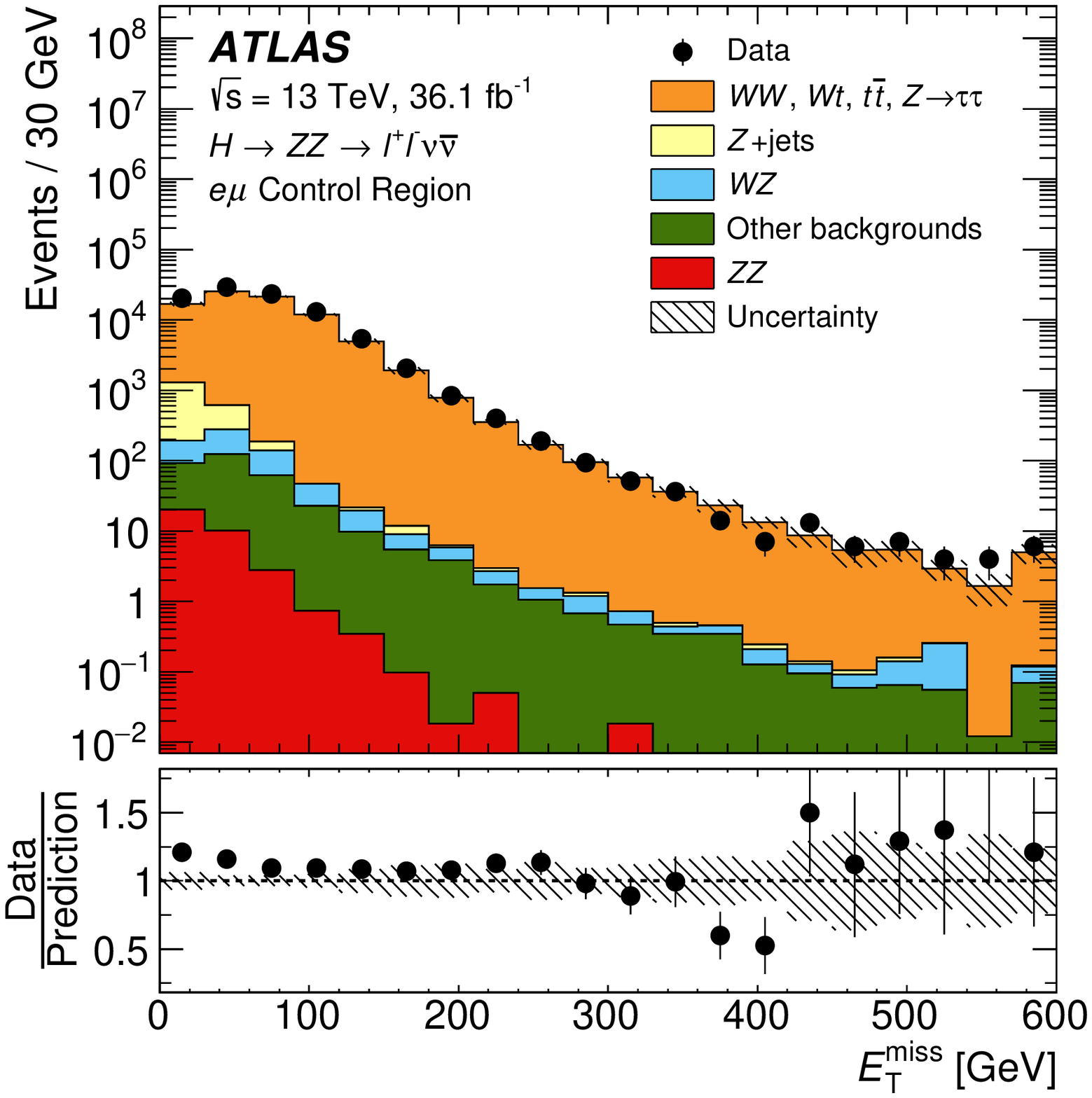}}
  \caption{Missing transverse momentum \met distribution \subref{fig:llvv_crs_sub1} for events in the \threel control region as defined in the text and \subref{fig:llvv_crs_sub2} for \emu lepton pairs after applying the dilepton invariant mass requirement, before applying the rest of the control region selection. The backgrounds are determined following the description in Section~\ref{sec:sigbkgllvv} and the last bin includes the overflow. The small excess below 120~\gev\ in \subref{fig:llvv_crs_sub2} arises from \Zjets\ background which is here taken from simulation, and lies outside the  control region. The error bars on the data points indicate the statistical uncertainty, while the systematic uncertainty in the prediction is shown by the \systband\ band. The lower panels show the ratio of data to prediction.}
\label{fig:llvv_crs}
\end{figure}

The \WZ background is modelled using simulation but a correction factor for its normalisation 
is extracted as the ratio of data to simulated events in a dedicated control region, 
after subtracting from data the non-\WZ background contributions.
The \WZ-enriched control sample, called the \threel control region,
is built by selecting \Zll candidates with an additional electron or muon.
This additional lepton is required to satisfy all selection criteria used for the other two leptons, 
with the only difference that its transverse momentum is required to be greater than 7~\gev.
The contamination from \Zjets and \ttbar events is reduced by vetoing events with at least one $b$-tagged jet
and by requiring the transverse mass of the \W boson (\mtw), 
built using the additional lepton and the \met vector, to be greater than 60~\gev.
The distribution of the missing transverse momentum for data and simulated
events in the \threel control region is shown in Figure~\ref{fig:llvv_crs_sub1}. 
The correction factor derived in the \threel control region is found to be $1.29\pm0.09$, 
where the uncertainty includes effects from the number of events in the control region 
as well as from experimental systematic uncertainties. Since there are few events after applying all the VBF selection requirements to the \WZ-enriched control sample, 
the estimation for the VBF-enriched  category is performed by including in the \threel control region 
only the requirement of at least two jets with $\pt > 30$~\gev. 
Finally, a transfer factor is derived from MC simulation by calculating 
the probability of events satisfying all analysis selection criteria and containing two jets with $\pt > 30$~\gev\ 
to satisfy the $\detajj > 4.4$ and $\mjj > 550$~\gev\ requirements. 
 An additional systematic uncertainty obtained from the comparison of the $\detajj$ distribution between \SHERPA and \POWHEGBOX generators is included to cover potential mismodellings of the VBF selection. Such systematic uncertainty is included in all background estimations   when extrapolating from a control region.

The non-resonant background includes mainly \WW, \ttbar and \Wt processes, but also \Ztt events in which
the $\tau$ leptons produce light leptons and \met. It is estimated by using  
a control sample of events with lepton pairs of different flavour (\emu), satisfying all analysis selection criteria.

Figure~\ref{fig:llvv_crs_sub2} shows the missing-transverse-momentum distribution for \emu events 
in data and simulation after applying the dilepton invariant-mass selection but before applying the other selection requirements. The non-resonant background in the \ee and \mm channels is estimated by applying
a scale factor ($f$) to the selected events in the \emu control region, such that:

\begin{equation*}
  \neebkg = \frac{1}{2} \times \nemu \times f, 
  \hspace{1cm} 
  \nmmbkg = \frac{1}{2} \times \nemu \times \frac{1}{f} ,
\label{eq::emu_cr}
\end{equation*}

where \neebkg and \nmmbkg are the numbers of electron- and muon-pair events 
estimated in the signal region and \nemu is the number of events 
in the $e^{\pm}\mu^{\mp}$ control sample with \ZZ, \WZ and other small backgrounds subtracted using simulation.
The factor $f$ takes into account the different selection efficiencies of \ee and \mm pairs at the level of the \Zll selection, 
and is measured from data as $f^2 = \needata/\nmmdata$, where \needata
and \nmmdata are the numbers of events passing the \Z boson mass requirement 
($76<m_{\ell\ell}<106$~\gev) in the electron and muon channel respectively.
As no events survive in the $e^{\pm}\mu^{\mp}$ control region after applying the full VBF selection, 
the background estimation is performed by including only the requirement of at least two jets with $\pt > 30$~\gev.
The efficiency of the remaining selection requirements on \detajj and \mjj is obtained from simulated events.

The number of \Zjets background events in the signal region is estimated from data, 
using a so-called ABCD method~\cite{STDM-2012-07}, since events with no genuine \met 
in the final state are difficult to model using simulation.
The method combines the selection requirements presented in Section~\ref{sec:selllvv} (with $n_{b\csthyphen \textrm{tags}}$ representing the number of $b$-tagged jets in the event) into two Boolean discriminants, 
\varone and \vartwo, defined as:

\begin{gather*}
  \varone \equiv \met > 120~\gev \andand \met / \HT > 0.4, \\ 
  \vartwo \equiv \fracpt < 0.2 \andand \Delta\phi(\ell\ell,\vecmet) > 2.7 \andand \Delta R_{\ell\ell} < 1.8 \andand n_{b\csthyphen \textrm{tags}} = 0,
\label{eq::abcd}
\end{gather*}
with all events required to pass the trigger and dilepton invariant-mass selections. 
The signal region (A) is thus obtained by requiring both \varone and \vartwo to be true, 
control regions B and C require only one of the two Boolean discriminants to be false (\varone and \vartwo respectively) and 
finally control region D is defined by requiring both \varone and \vartwo to be false.
With this definition, 
an estimate of the number of events in region A is given 
by $\nnn{A}{est} = \nnn{C}{obs} \times (\nnn{B}{obs}/\nnn{D}{obs})$, where \nnn{X}{obs}
is the number of events observed in region X after subtracting non-\Zboson-boson backgrounds.
This relation holds as long as the correlation between \varone and \vartwo is small, 
and this is achieved by introducing two additional requirements 
on control regions B and D, namely \met > 30~\gev\ and \met / \HT > 0.1. The estimation of the \Zjets background was cross-checked with another approach in which a control region is defined 
by inverting the analysis selection on $\met / \HT$ and then using \Zjets\ MC simulation 
to perform the extrapolation to the signal region, yielding results compatible with the ABCD method. 
Finally, the estimate for the VBF-enriched category is performed by extrapolating the inclusive result obtained 
with the ABCD method to the VBF signal region, extracting the efficiency of the two-jet, \detajj and \mjj selection criteria from \Zjets\ simulation.

The \Wjets and multi-jet background contributions are estimated from data using a so-called fake-factor method~\cite{STDM-2011-16}.
A control region enriched in fake leptons or non-prompt leptons from decays of hadrons is designed by requiring one lepton to pass 
all analysis requirements (baseline selection) and the other one to not pass either the lepton ``medium'' 
identification or the isolation criteria (inverted selection).
The background in the signal region is then derived using a transfer factor, measured in a data sample enriched in \Zjets\ events,
as the ratio of jets passing the baseline selection to those passing the inverted selection.

Finally, the background from the \ttv and \vvv processes is estimated using MC simulation.
\subsection{Signal and background modelling}
\label{sec:h2l2v_sig_model}
The modelling of the transverse mass \mt distribution for signal and background
is based on templates derived from fully-simulated events and afterwards used to fit the data.
In the case of a narrow resonance, simulated MC events generated for fixed mass hypotheses 
as described in Section~\ref{sec:samples} are used as the inputs in the 
moment-morphing technique~\cite{Baak:2014fta} to obtain the \mt distribution for 
any other mass hypothesis. 

The extraction of the interference terms for the LWA case is performed in the same way as in the \llll final state, as described in Section~\ref{sec:h4l_sig_model}. In the case of the \llvv final state a correction factor, extracted as a function of $m_{ZZ}$, is used 
to reweight the interference distributions obtained at particle level to account for reconstruction 
effects. The final expected LWA \mt distribution is obtained from the combination of the interference 
distributions with simulated \mt distributions, which are interpolated between the simulated mass 
points with a weighting technique using the Higgs propagator, a method similar to that used for the interference.

\section{Systematic uncertainties}
\label{sec:syst}
The systematic uncertainties can be classified into experimental and theoretical uncertainties. 
The first category relates to the reconstruction and 
identification of leptons and jets, their energy scale and resolution, and the integrated luminosity.
Systematic uncertainties in the data-driven background estimates are also included in this category.
The second category includes uncertainties in the theoretical description of the signal and background processes.

In both cases the uncertainties are implemented as additional nuisance parameters (NP) 
that are  constrained by a Gaussian distribution in the profile
likelihood ratio, as discussed in Section \ref{sec:stats}. 
The uncertainties affect the signal acceptance, its selection efficiency and the discriminant distributions as well as the background estimates for both final states. 
Each source of uncertainty is either fully correlated or anti-correlated among the different channels and categories.

\subsection{Experimental uncertainties}
The uncertainty in the combined 2015 and 2016 integrated luminosity is \lumiErr. 
This is derived from a preliminary calibration of the luminosity scale using $x$--$y$ beam-separation scans performed in August 2015 and May 2016, 
following a methodology similar to that detailed in Ref.~\cite{lumi2016}.
 
The lepton identification and reconstruction efficiency and energy/momentum scale and resolution
are derived from data using large samples of $J/\psi \rightarrow \ell\ell$  and $Z \rightarrow \ell\ell$ decays.
The uncertainties in the reconstruction performance are computed following the method described in Ref.~\cite{PERF-2015-10} for muons 
and Ref.~\cite{PERF-2016-01} for electrons.
Typical uncertainties in the identification and reconstruction efficiency are in the range 0.5\%--3.0\% for muons 
and 1.0\%--1.7\% for electrons. 
The uncertainties in the electron energy scale, the muon momentum scale and their resolutions are small,  and are fully correlated between the two searches (\llll and \llvv final states).

The uncertainties in the jet energy scale and resolution have several sources, including
uncertainties in the absolute and relative \textit{in situ} calibration, the correction for \pu, 
the flavour composition and response~\cite{PERF-2016-04}. 
These uncertainties are separated into independent components, which are fully correlated between the two searches.   
They vary from 4.5\% for jets with transverse momentum \pt = 20~\gev, 
decreasing to 1\% for jets with $\pt = 100$--$1500$~\gev\ and increasing again to 3\% for jets with higher \pt, 
for the average \pu conditions of the 2015 and 2016 data-taking period.
 
Uncertainties in the lepton and jet energy scales are propagated to the uncertainty in the \met. 
Additionally, the uncertainties from the momentum scale and resolution of the tracks that are not 
associated with any identified lepton or jet contribute 8\% and 3\% respectively, to the uncertainty in the \met value.

The efficiency of the lepton triggers in events with reconstructed leptons is nearly 100\%, 
and hence the related uncertainties are negligible. 

\subsection{Theoretical uncertainties}
For simulated signal and backgrounds, theoretical modelling uncertainties associated with
the PDFs, missing QCD higher-order corrections (via variations of factorisation and renormalisation scales), 
and parton showering are considered.

For all signal hypotheses under consideration, the largest theoretical modelling uncertainties are due to
missing QCD higher-order corrections and parton showering. 
The missing QCD higher-order corrections for ggF production events that fall into the
VBF-enriched category are accounted for by varying the scales in \MGMCatNLO and affect the signal acceptance by 10\%.  
Parton showering uncertainties are of order 10\% and are estimated by comparing \PYTHIAV{8.212} to \progname{Herwig++}~\cite{Bahr:2008pv}.  

For the \qqZZ background, the effect of the PDF uncertainties in the full mass range 
varies between 2\% and 5\% in all categories, and that of missing QCD higher-order corrections is about 10\% 
in the ggF-enriched categories and 30\% in the VBF-enriched category.
The parton-shower uncertainties result in less than 1\% impact in the ggF-enriched categories and about 10\%
impact in the VBF-enriched category.

For the \ggZZ background, as described in Section~\ref{sec:samples}, a 60\% relative uncertainty in the inclusive cross section is considered, while a 100\% uncertainty is assigned in the VBF-enriched category.

\section{Results and interpretations}
\label{sec:results}
\subsection{Statistical procedure}
\label{sec:stats}

The statistical treatment of the data follows the procedure for the Higgs-boson search combination~\cite{ATL-PHYS-PUB-2011-011, HIGG-2012-17},
and is implemented with RooFit~\cite{roofit:2003} and RooStats~\cite{roostats:2010}.
The test statistic employed for hypothesis testing and limit setting is the profiled likelihood ratio $\Lambda(\alpha, \boldsymbol{\theta})$,
which depends on one or more parameters of interest $\alpha$, and additional nuisance parameters $\boldsymbol{\theta}$.
The parameter of interest is the cross section times branching ratio for heavy-resonance production, assumed to be correlated between the two searches.
The nuisance parameters represent the estimates of the systematic uncertainties and are each constrained by a Gaussian distribution.
For each category of each search, a likelihood fit to the kinematic distribution of a discriminating variable is used to further separate signal from background.
The $\llll$ final state uses \mfl as the discriminant in each category, while the $\llvv$ final state uses \mt in each category except for the VBF-enriched one where only the overall event counts are used.

As discussed in Section~\ref{sec:syst}, the signal acceptance uncertainties, and many of the background theoretical and experimental uncertainties, are treated as fully correlated between the searches. A given correlated uncertainty is modelled in the fit by using a nuisance parameter common to all of the searches.
The impact of a systematic uncertainty on the result depends on the production mode and the mass hypothesis.
For ggF production, at lower masses the luminosity uncertainty, the modelling uncertainty of the \Zjets\ background and the statistical uncertainty in the $e\mu$ control region of the \llvv final state dominate,
and at higher masses the uncertainties in the electron-isolation efficiency become important, as also seen in VBF production.
For VBF production, the dominant uncertainties come from the theoretical predictions of the \ZZ events in the VBF category.
Additionally at lower masses, the \pu\ reweighting and the jet-energy-resolution uncertainties are also important.
Table~\ref{tab:ranking} shows the impact of the leading systematic uncertainties on the predicted signal event yield when the cross section times branching ratio is set to the expected upper limit (shown in Figure~\ref{fig:NWAlimits}), for ggF and VBF production modes. The impact of the uncertainty in the integrated luminosity, 3.2\%, enters both in
the normalisation of the fitted number of signal events as well as in the background predicted 
by simulation.  This leads to a luminosity uncertainty which varies from 4\% to 7\% across the
mass distribution, depending on the signal-to-background ratio.

\begin{table}[htbp!]
  \caption{\label{tab:ranking}
  Impact of the leading systematic uncertainties on the 
  predicted signal event yield which is set to the expected upper limit, expressed as a percentage of the yield 
   for the ggF (left) and VBF (right) production modes at $\mres = 300$, 600, and $1000~\gev$.
  }
  \small
  \centering
  \begin{tabular}{%
   l
   S[table-format=2.1, table-number-alignment=right,
     round-mode=figures, round-precision=2] 
   |
   l
   S[table-format=2.1, table-number-alignment=right,
     round-mode=figures, round-precision=2]
  }
    \toprule
    \multicolumn{2}{c|}{ggF production} &  \multicolumn{2}{c}{VBF production} \\
    \multicolumn{2}{l|}{Systematic source    \hfill Impact [\%]}   & \multicolumn{2}{l}{Systematic source \hfill Impact [\%]}\\
    \midrule
    \multicolumn{4}{c}{$\bigstrut$  $\mres=300~\gev$}\\
    \midrule
    \sysLumi & 4 & \shw & 9 \\
    \abcd & 3.3 & \jes & 4 \\
    \shw & 3.2 & \sysLumi & 4 \\
    \emuemu & 3.2 & \qcdvvVBF & 4 \\
    \midrule
    \multicolumn{4}{c}{$\bigstrut$  $\mres=600~\gev$}\\
    \midrule
    \sysLumi & 6 & \shw & 6 \\
    \prw & 5 & \prw & 6 \\
    \abcd & 4 & \jes & 6 \\
    \qcdvv & 3.1 & \sysLumi & 4 \\
    \midrule
    \multicolumn{4}{c}{$\bigstrut$  $\mres=1000~\gev$}\\
    \midrule
      \sysLumi & 4 & \shw & 6 \\
      \qcdggvv & 2.3 & \jes & 5 \\
      \jvt & 1.9 & \abcd & 4 \\
      \abcd & 1.8  & \sysLumi & 4 \\
    \bottomrule
  \end{tabular}
\end{table}

\subsection{General results}
\label{sec:resul}

\begin{figure}[tbp]
  \subfigure[\label{fig:m4l_sig_sub1}]{\includegraphics[width=0.49\textwidth]{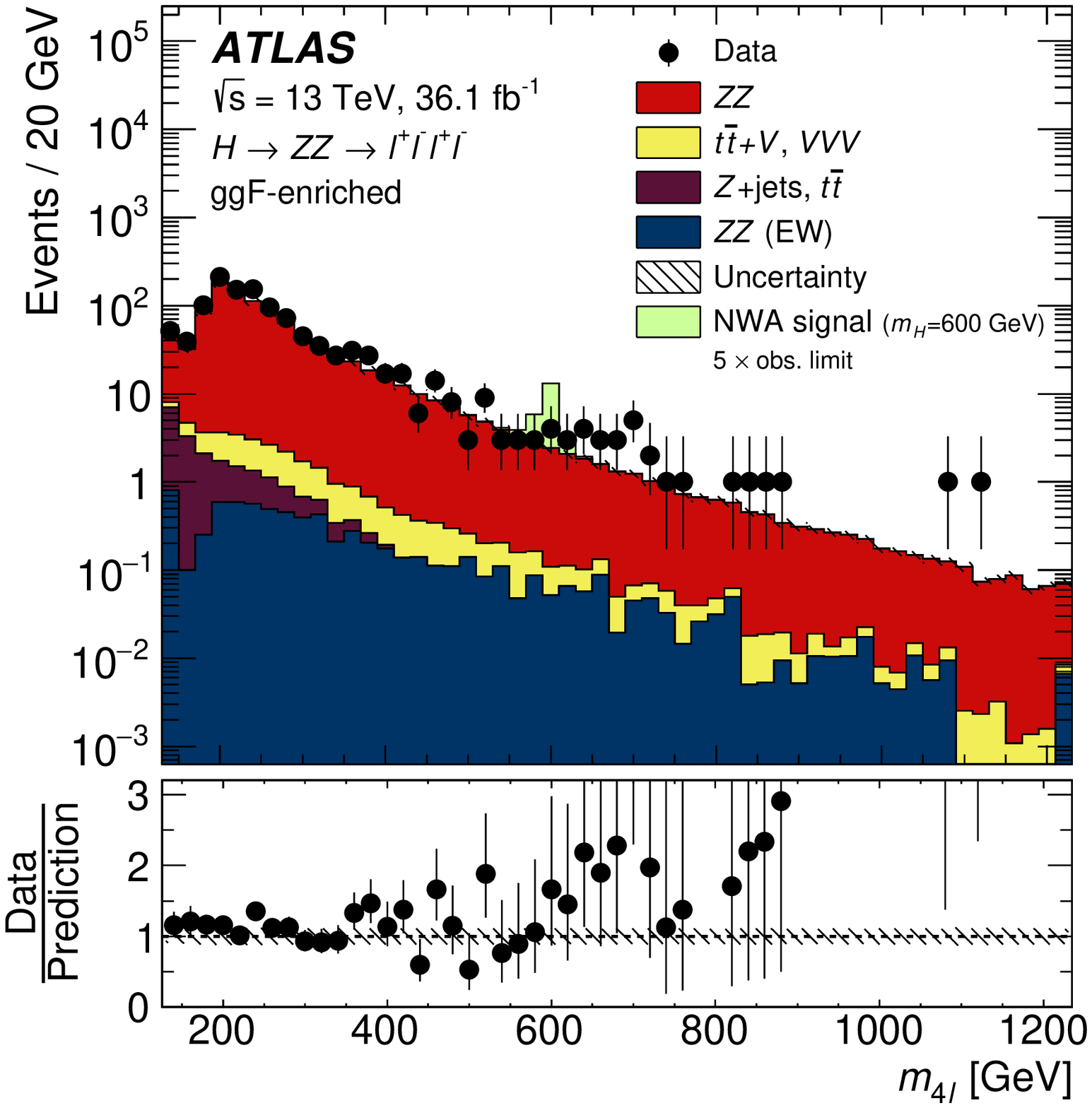}}
  \subfigure[\label{fig:m4l_sig_sub2}]{\includegraphics[width=0.49\textwidth]{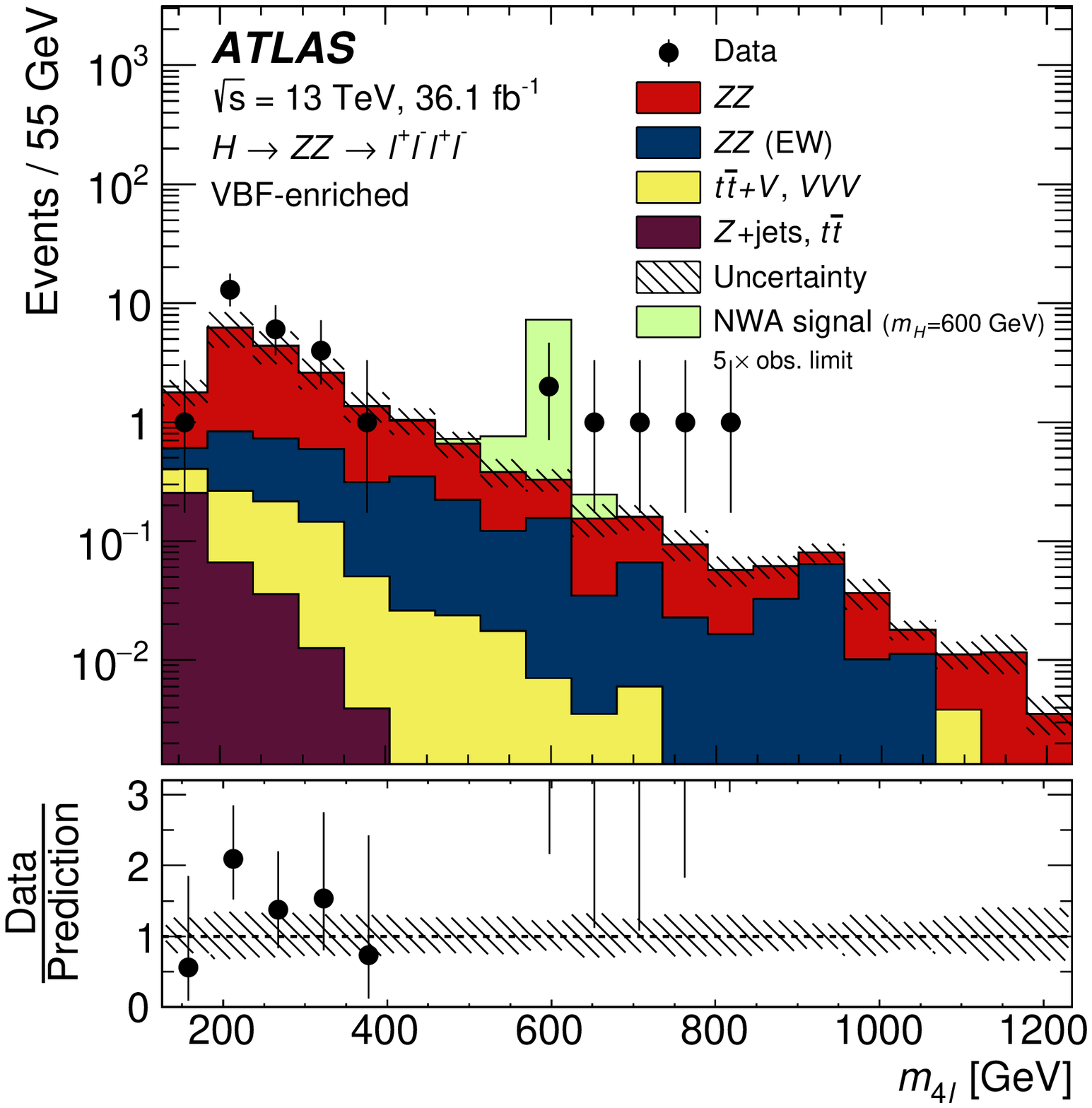}}
  \caption{Distribution of the four-lepton invariant mass \mfl in the $\llll$ search for \subref{fig:m4l_sig_sub1} the ggF-enriched category and \subref{fig:m4l_sig_sub2} the VBF-enriched category. The backgrounds are determined following the description in Section~\ref{sec:bkg4l} and the last bin includes the overflow. The simulated $\mres = 600$~\gev\ signal is normalized to a cross section corresponding to five times the observed limit given in Section~\ref{sec:res_NWA}. The error bars on the data points indicate the statistical uncertainty, while the systematic uncertainty in the prediction is shown by the \systband\ band. The lower panels show the ratio of data to prediction.  }   
\label{fig:m4l_sig} 
\end{figure}

The numbers of observed candidate events with mass above 130~\gev\ 
together with the expected background yields are presented in Table~\ref{tab:H4lyields} for each of the four categories of the $\llll$ analysis. The \mfl spectrum for the ggF-enriched and VBF-enriched categories is shown in Figure~\ref{fig:m4l_sig}. 

\begin{figure}[tbp]
  \subfigure[\label{fig:llvv_sig_sub1}]{\includegraphics[width=0.49\textwidth]{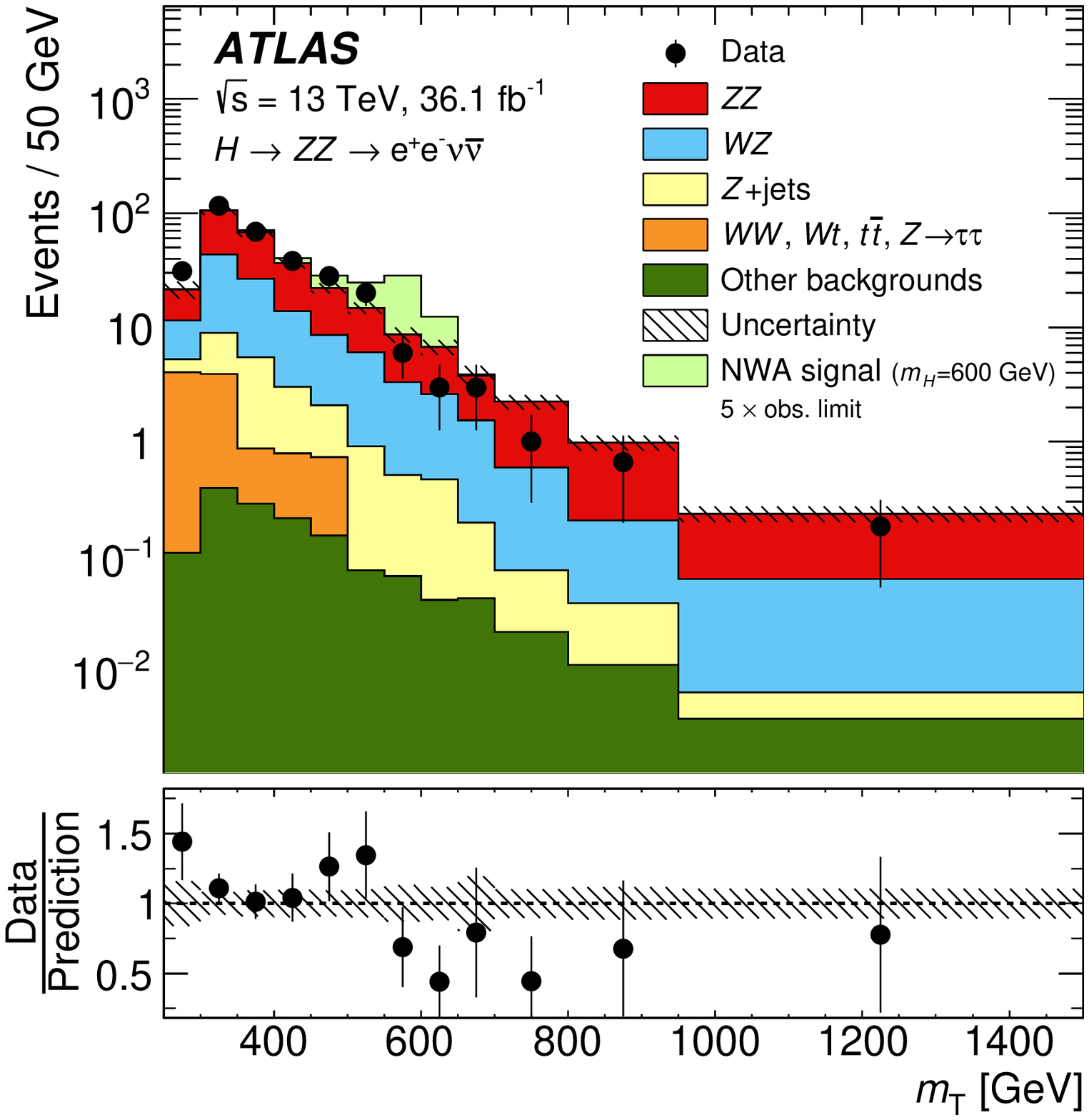}}
  \subfigure[\label{fig:llvv_sig_sub2}]{\includegraphics[width=0.49\textwidth]{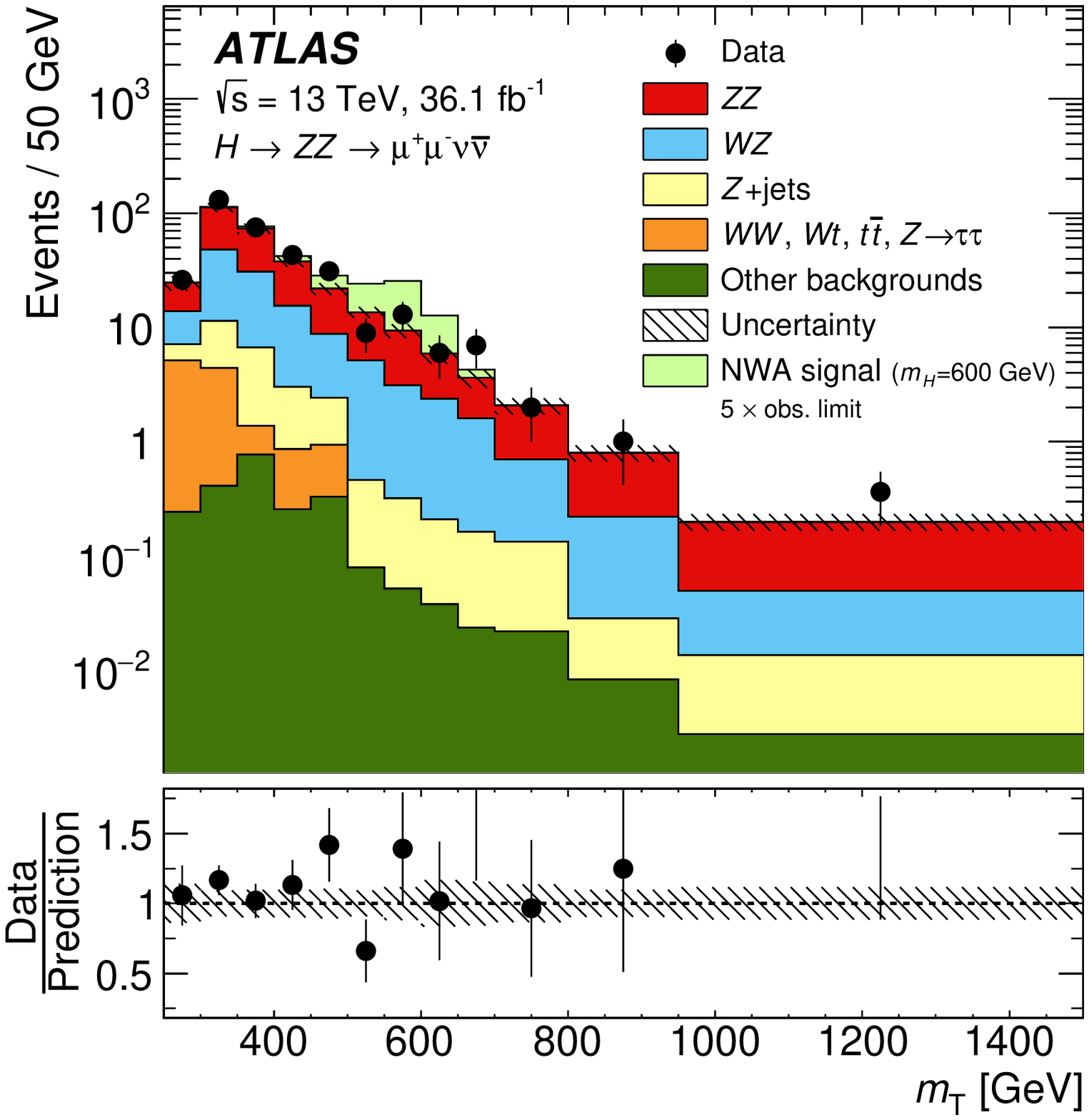}}
  \caption{Transverse mass \mt{} distribution in the $\llvv$ search for \subref{fig:llvv_sig_sub1} the electron channel and \subref{fig:llvv_sig_sub2} the muon channel, including events from both the ggF-enriched and the VBF-enriched categories. The backgrounds are determined following the description in Section~\ref{sec:sigbkgllvv} and the last bin includes the overflow. The simulated $\mres = 600$~\gev\ signal is normalized to a cross section corresponding to five times the observed limit given in Section~\ref{sec:res_NWA}. The error bars on the data points indicate the statistical uncertainty and markers are drawn at the bin centre. The systematic uncertainty in  the prediction is shown by the \systband\ band. The lower panels show the ratio of data to prediction.  }  \label{fig:llvv_sig}
\end{figure}

Table~\ref{tab:ZZllnn_yields_postfit} contains the number of observed candidate events along with
the background yields for the $\llvv$ analysis, while Figure~\ref{fig:llvv_sig} shows the \mt
distribution for the electron and muon channels with the ggF-enriched and VBF-enriched categories 
combined.

\begin{table}[tbp]
  \caption{\label{tab:H4lyields}
    \llll search: expected and observed numbers of events for \mfl $> 130$~\gev, together with their statistical and systematic uncertainties, 
    for the ggF- and VBF-enriched categories.
  }
  \centering
  \tiny
  \resizebox{\textwidth}{!}{
    \begin{tabular}{l r@{ $\pm$ }c@{ $\pm$ }l r@{ $\pm$ }c@{ $\pm$ }l r@{ $\pm$ }c@{ $\pm$ }l r@{ $\pm$ }c@{ $\pm$ }l }
      \toprule
      \multirow{2}{*}{Process} & \multicolumn{9}{c}{ggF-enriched categories}  & \multicolumn{3}{c}{\multirow{2}{*}{VBF-enriched category}} \\
      & \multicolumn{3}{c}{$4\mu$ channel} & \multicolumn{3}{c}{$2e2\mu$ channel} & \multicolumn{3}{c}{$4e$ channel} & \multicolumn{3}{c}{} \\
      \midrule
      \ZZ                 & 297  & 1    & 40   & 480  & 1    & 60   & 193  & 1    & 25   & 15   & 0.1  & 6.0 \\ 
      \ZZ (EW)            & 1.92 & 0.11 & 0.19 & 3.36 & 0.14 & 0.33 & 1.88 & 0.12 & 0.20 & 3.0  & 0.1  & 2.2 \\
      \Zjets/\ttbar/\WZ   & 3.7  & 0.1  & 0.8  & 7.8  & 0.1  & 1.1  & 4.4  & 0.1  & 0.8  & 0.37 & 0.01 & 0.05 \\
      Other backgrounds   & 5.1  & 0.1  & 0.6  & 8.7  & 0.1  & 1.0  & 4.0  & 0.1  & 0.5  & 0.80 & 0.02 & 0.30 \\
      \midrule                                                      
      Total background    & 308  & 1    & 40   & 500  & 1    & 60   & 203  & 1   & 25   & 19.5 & 0.2  & 8.0  \\
      \midrule
      Observed                 &\multicolumn{3}{c}{357} & \multicolumn{3}{c}{545}  & \multicolumn{3}{c}{256}  & \multicolumn{3}{c}{31}  \\
      \bottomrule
    \end{tabular}
  }
\end{table}

\begin{table}[tbp]
\caption{\label{tab:ZZllnn_yields_postfit}
\llvv search: expected and observed number of events together with their statistical and systematic uncertainties, 
for the ggF- and VBF-enriched categories.
}
\centering
\begin{tabular}{l r@{ $\pm$ }c@{ $\pm$ }l r@{ $\pm$ }c@{ $\pm$ }l r@{ $\pm$ }c@{ $\pm$ }l }
\toprule
\multirow{2}{*}{Process} & \multicolumn{6}{c}{ggF-enriched categories}  & \multicolumn{3}{c}{\multirow{2}{*}{VBF-enriched category}} \\
& \multicolumn{3}{c}{\ee channel} & \multicolumn{3}{c}{\mm channel} & \multicolumn{3}{c}{} \\
\midrule
 \ZZ                      & 177  & 3    & 21 & 180     & 3    & 21   & 2.1  & 0.2  & 0.7 \\
 \WZ                      & 93   & 2    & 4    & 99.5  & 2.3  & 3.2  & 1.29 & 0.04 & 0.27 \\
 \WW/\ttbar/\Wt/\Ztt      & 9.2  & 2.2  & 1.4  & 10.7  & 2.5  & 0.9  & 0.39 & 0.24 & 0.26 \\
 \Zjets                   & 17   & 1    & 11   & 19    & 1    & 17   & 0.8  & 0.1  & 0.5 \\
 Other backgrounds        & 1.12 & 0.04 & 0.08 & 1.03  & 0.04 & 0.08 & 0.03 & 0.01 & 0.01 \\
\midrule                                                      
 Total background         & 297  & 4    & 24   & 311   & 5    & 27   & 4.6  & 0.4  & 0.9 \\ 
\midrule
 Observed                 &\multicolumn{3}{c}{320} & \multicolumn{3}{c}{352}  & \multicolumn{3}{c}{9}  \\
\bottomrule
\end{tabular}
\end{table}

In the \llll search, two excesses are observed in the data for \mfl around 240 and 700~\gev,
each with a local significance  of $3.6\sigma$ estimated in the asymptotic approximation, 
assuming the signal comes only from ggF production. 
The global significance is $2.2\sigma$ and is  calculated, for each excess individually, using the NWA, in the range of 200~\gev $<$ \mres $<$ 1200~\gev\ using pseudo-experiments.
 
The excess at 240~\gev\ is observed mostly in the $4e$ channel, while the one at 700~\gev\ is
observed in all channels and categories.  No significant deviation from the expected background
is observed in the \llvv final state. The excess observed in the \llll search
at a mass around 700~\gev\ is excluded at 95\% confidence level (CL) by the \llvv search, which is more
sensitive in this mass range. The excess at 240~\gev\ is not covered by the \llvv search, the
sensitivity of which starts from 300~\gev. When combining the results from the two final
states, the largest deviation with respect to the background expectation is observed around
700~\gev\ with a  global significance of less than
$1\sigma$ and a local significance of about $2\sigma$. 
The combined yield of the two final states is 1870 events observed in data compared to
$1643\pm164$ (combined statistical and systematic uncertainty) for the expected background. This
corresponds to a $1.3\sigma$ global excess in data.  
Since no significant excess is found, the results are interpreted as upper
limits on the production cross section of a spin-0 or spin-2 resonance.

\subsection{Spin-0 resonance interpretation}
\label{sec:resul}
Limits from the combination of the two searches in the context of a spin-0 resonance are described below. 

\subsubsection{NWA interpretation}
\label{sec:res_NWA}

\begin{figure}[tbp]
  \subfigure[\label{fig:NWAlimits_sub1}]{\includegraphics[width=0.49\textwidth]{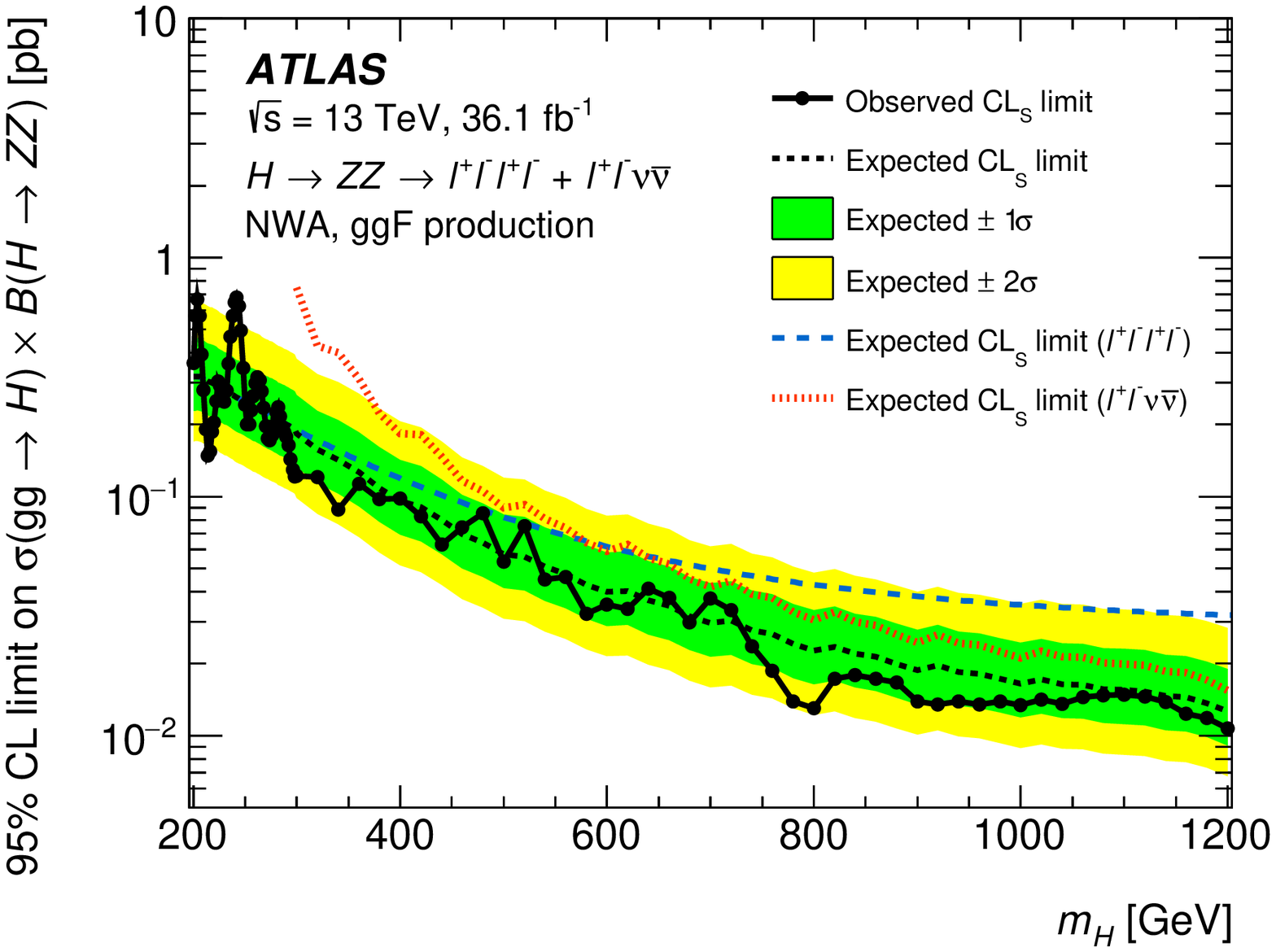}}
  \subfigure[\label{fig:NWAlimits_sub2}]{\includegraphics[width=0.49\textwidth]{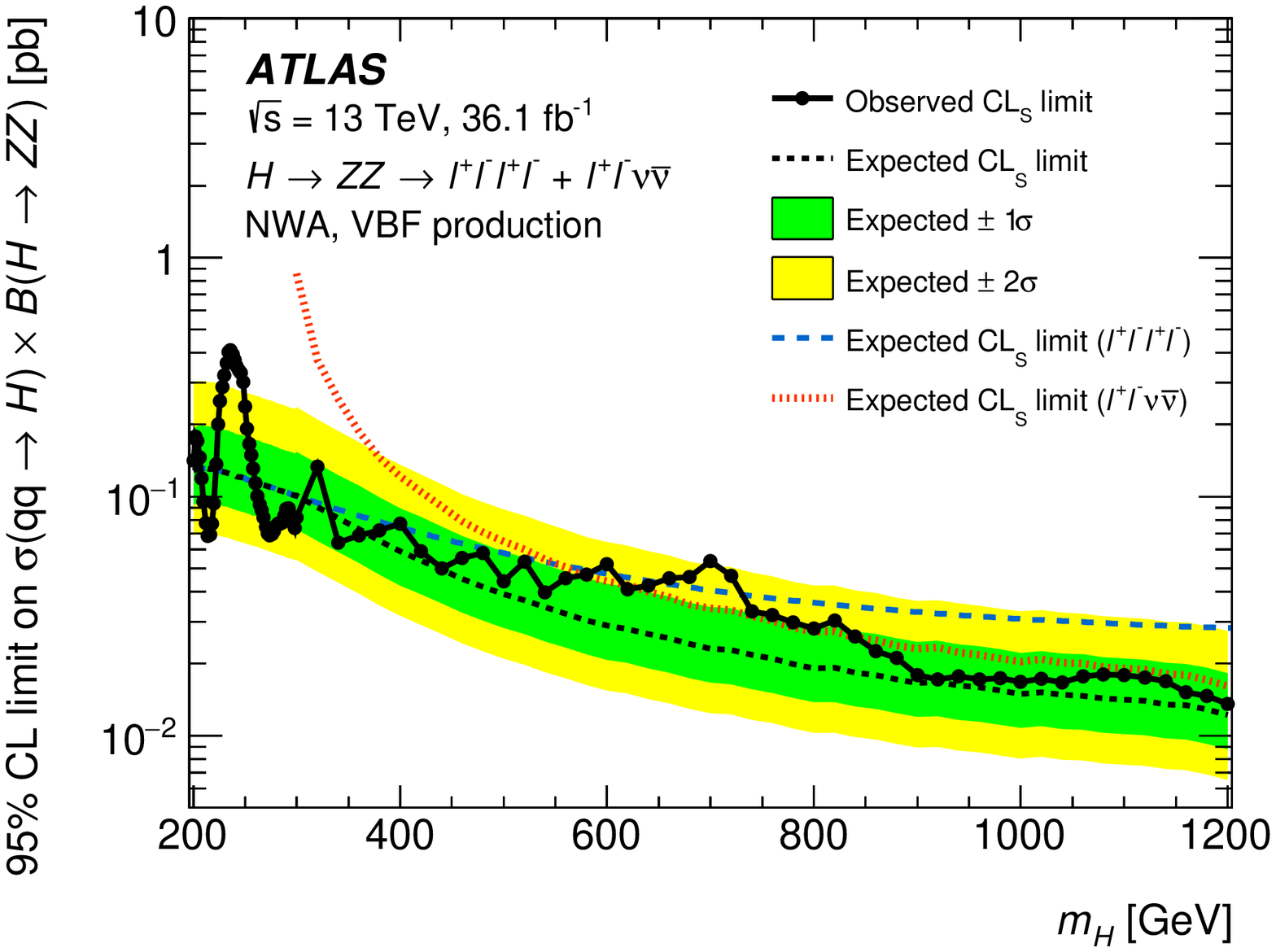}}
  \caption{The upper limits at 95\% CL  on  the cross section times branching ratio as a function of the heavy resonance mass $\mres$ for \subref{fig:NWAlimits_sub1} the ggF production mode($\xsggF\times\mbr(\htozz)$) and \subref{fig:NWAlimits_sub2} for the VBF production mode  ($\xsVBF\times\mbr(\htozz)$)  in the case of the NWA. The green  and yellow  bands represent the $\pm$$1\sigma$ and $\pm$$2\sigma$ uncertainties in the expected limits. The dashed coloured lines indicate the expected limits obtained from the individual searches.
  \label{fig:NWAlimits}}
\end{figure}

Upper limits on the cross section times branching ratio ($\sigma \times\mbr(\htozz)$) for a 
heavy resonance are obtained as a
function of \mres\ with the $\mathrm{CL}_{\mathrm{s}}$ procedure~\cite{Read:2002hq} in the asymptotic approximation
from the combination of the two final states. It is assumed that an additional heavy scalar would be
produced predominantly via the ggF and VBF processes but that the ratio of the two production
mechanisms is unknown in the absence of a specific model. For this reason, fits for the ggF and VBF
production processes are done separately, and in each case the other process is allowed to float in
the fit as an additional nuisance parameter. Figure~\ref{fig:NWAlimits} presents the observed and
expected limits at 95\% CL on $\sigma \times\mbr(\htozz)$ of a narrow scalar resonance for
the ggF (left) and VBF (right) production modes, as well as the expected limits from the $\llll$ and
$\llvv$ searches. This result is valid for models in which the width is less than 0.5\% of \mres.
When combining the two final states, the 95\% CL upper limits range from 0.68~pb at $\mres = 242$~\gev\ 
to 11~fb at $\mres = 1200$~\gev\ for the ggF production mode
and from 0.41~pb at $\mres = 236$~\gev\ to 13~fb at $\mres = 1200$~\gev\ 
for the vector-boson fusion production mode.  Compared with the results from \runa~\cite{HIGG-2013-20}, where all four final states 
of \ZZ decays were combined, the exclusion region presented here is significantly 
extended considering that the ratios of parton luminosities~\cite{Martin:2009iq} increase by factors of
about two to seven for heavy scalar masses from 200~\gev\ to 1200~\gev.

\subsubsection{LWA interpretation}
\label{sec:res_LWA}
\begin{figure}[tbp]
  \centering
  \subfigure[\label{fig:LWAlimits_1}]{\includegraphics[width=0.45\textwidth]{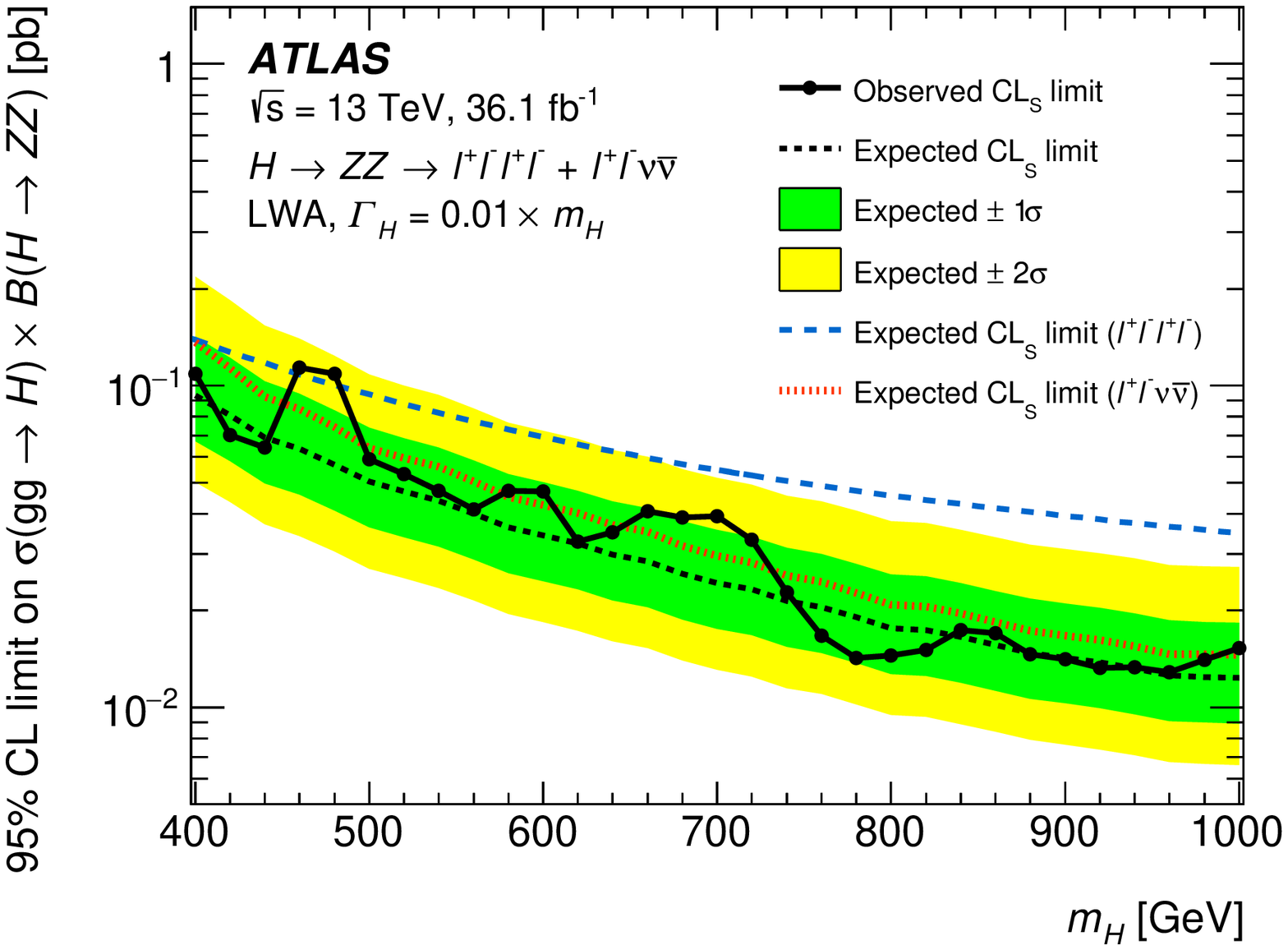}}
  \subfigure[\label{fig:LWAlimits_5}]{\includegraphics[width=0.45\textwidth]{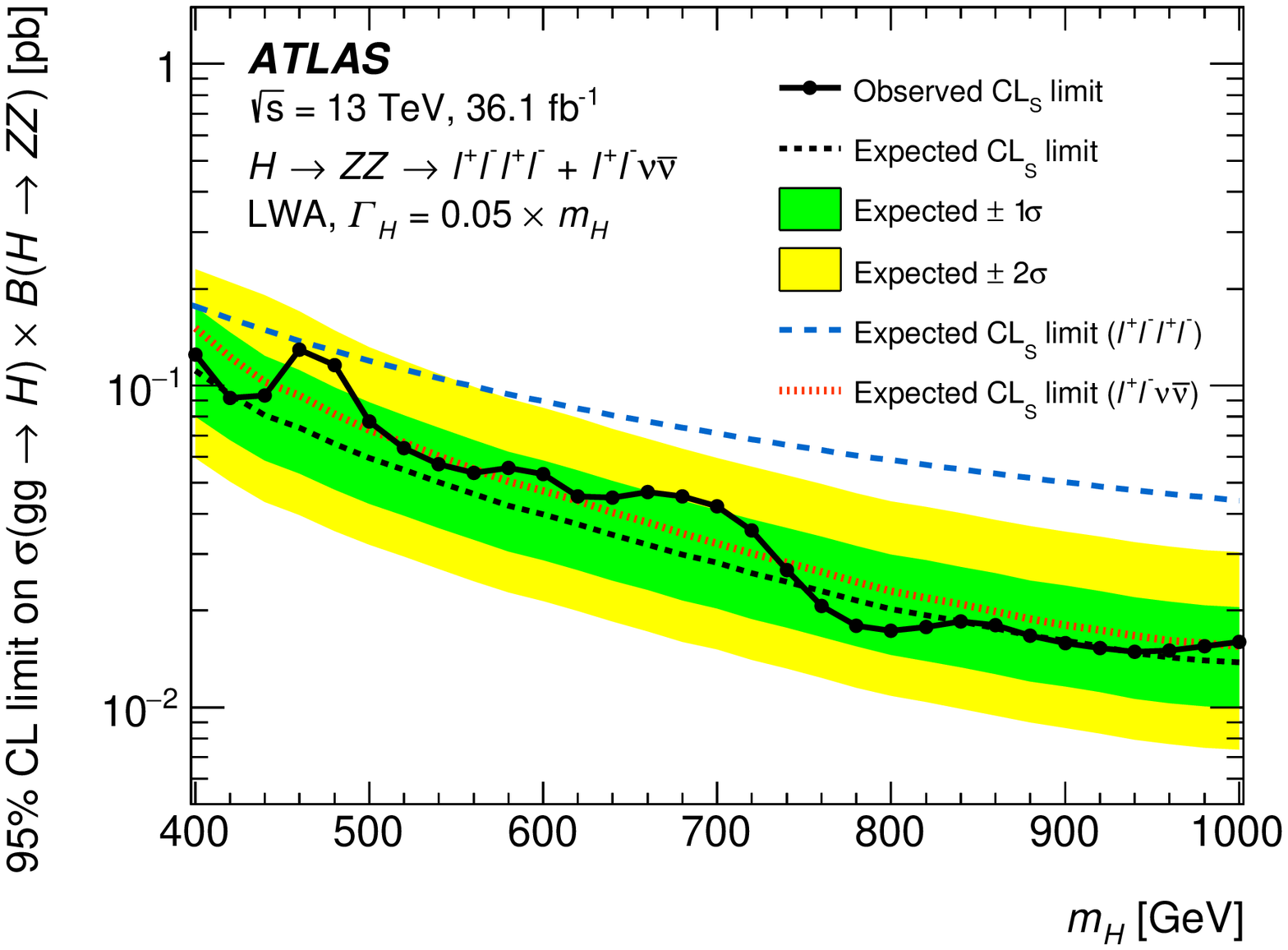}} 
  \subfigure[\label{fig:LWAlimits_10}]{\includegraphics[width=0.45\textwidth]{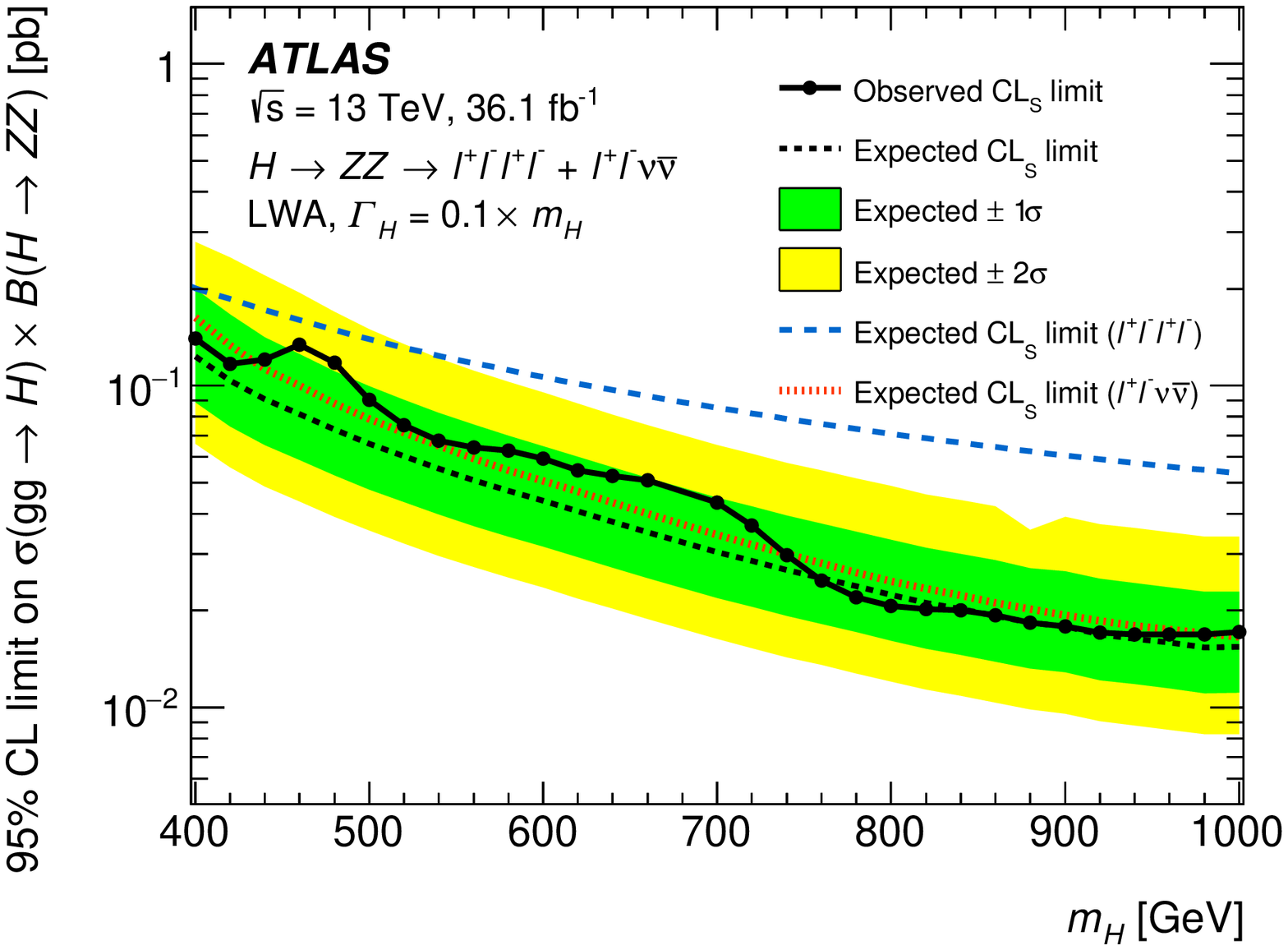}}
  \caption {The upper limits at 95\% CL on the cross section for the ggF production mode times branching ratio ($\sigma_{\textrm{ggF}}\times\mbr(\htozz)$) as function of \mres\ for an additional heavy scalar assuming a width of~\subref{fig:LWAlimits_1} 1\%,~\subref{fig:LWAlimits_5} 5\%, and~\subref{fig:LWAlimits_10} 10\% of \mres. The green  and yellow  bands represent the $\pm$$1\sigma$ and $\pm$$2\sigma$ uncertainties in the expected limits. The dashed coloured lines indicate the expected limits obtained from the individual searches. 
\label{fig:LWAlimits}}
\end{figure}

In the case of the LWA, limits on the cross section for the ggF production mode times branching ratio ($\sigma_{\textrm{ggF}} \times\mbr(\htozz)$) are set for different 
widths of the heavy scalar. The interference between the heavy scalar and the SM Higgs boson,
\IntHh, as well as the heavy scalar and the \ggZZ continuum, \IntHB, are modelled by either
analytical functions or reweighting the signal-only events as explained in
Sections~\ref{sec:h4l_sig_model} and~\ref{sec:h2l2v_sig_model}.  Figures~\ref{fig:LWAlimits_1},
\ref{fig:LWAlimits_5}, and \ref{fig:LWAlimits_10} show the limits for a width of 1\%, 5\% and 10\% of
\mres respectively. The limits are set for masses of \mres\ higher than 400~\gev.

\subsubsection{2HDM interpretation}
\label{sec:res_2HDM}
A search in the context of a CP-conserving 2HDM is also
presented. This model has five physical Higgs bosons after electroweak symmetry breaking: two CP-even, one CP-odd, and two charged. The model considered here has seven free parameters: the
Higgs boson masses, the ratio of the vacuum expectation values of the two doublets ($\tan\beta$), the
mixing angle between the CP-even Higgs bosons ($\alpha$), and the potential parameter $m_{12}^2$ that mixes the two Higgs doublets.
The two Higgs doublets $\Phi_1$ and $\Phi_2$ can couple to
leptons and up- and down-type quarks in several ways.  In the Type-I model,
$\Phi_2$ couples to all quarks and leptons, whereas for Type-II,
$\Phi_1$ couples to down-type quarks and leptons and $\Phi_2$
couples to up-type quarks.  The ``lepton-specific'' model is similar to Type-I except for the fact 
that the leptons couple to $\Phi_1$, instead of $\Phi_2$; the ``flipped'' model is similar to Type-II except that the leptons 
couple to $\Phi_2$, instead of $\Phi_1$. 
In all these models, the coupling of the heaviest CP-even Higgs boson to vector bosons is proportional to
$\cos(\beta-\alpha)$. In the limit $\cos(\beta-\alpha) \rightarrow 0$, the
light CP-even Higgs boson is indistinguishable from a SM
Higgs~boson with the same mass. In the context of \htozz\ decays there is no direct coupling of the
Higgs~boson to leptons, and so only the Type-I and -II interpretations are
presented.

\begin{figure}[tbp]
  \centering
  \subfigure[\label{fig:2HDM_tanb_cba_mH200_sub1}]{\includegraphics[width=0.49\textwidth]{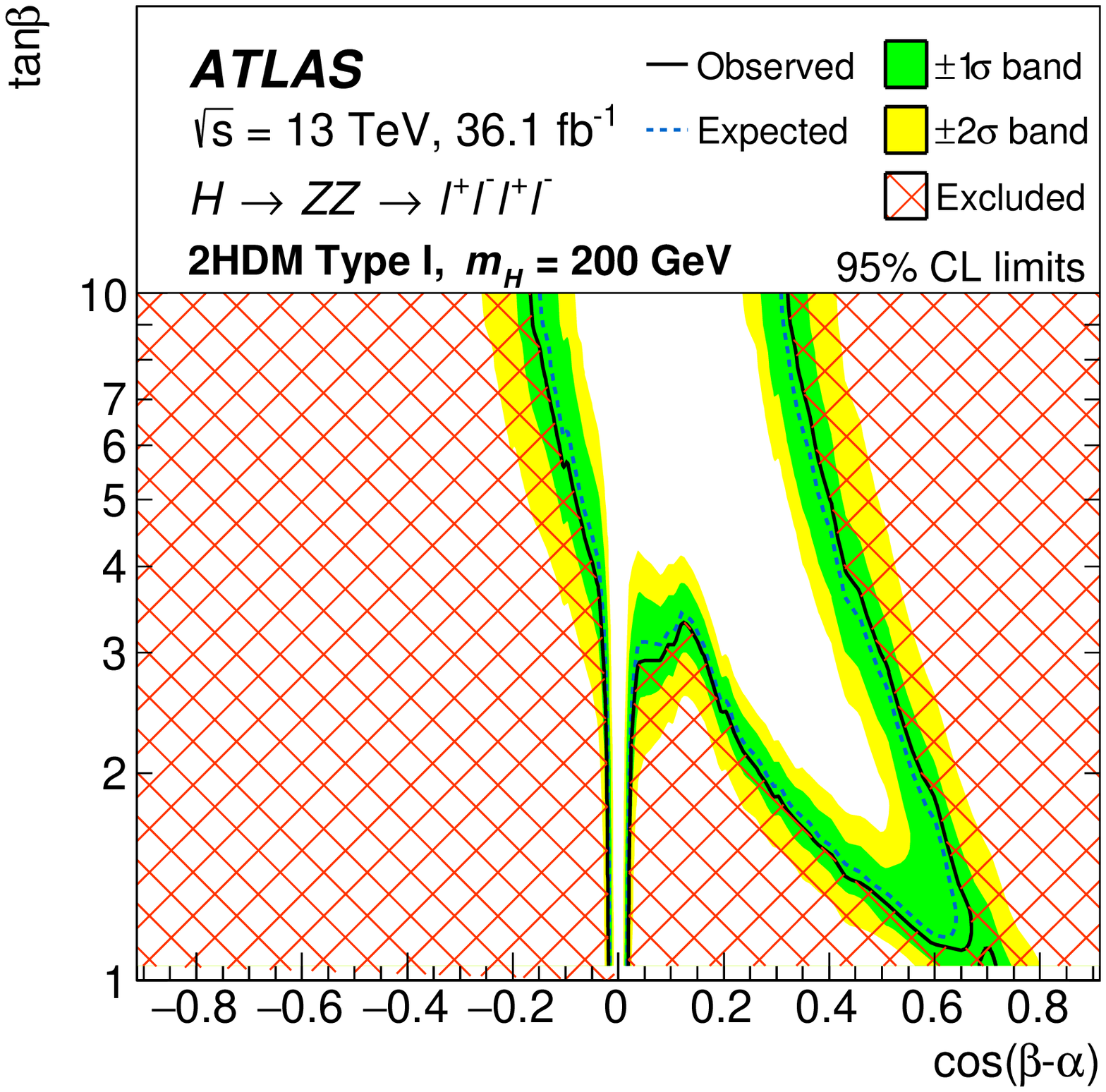}}
  \subfigure[\label{fig:2HDM_tanb_cba_mH200_sub2}]{\includegraphics[width=0.49\textwidth]{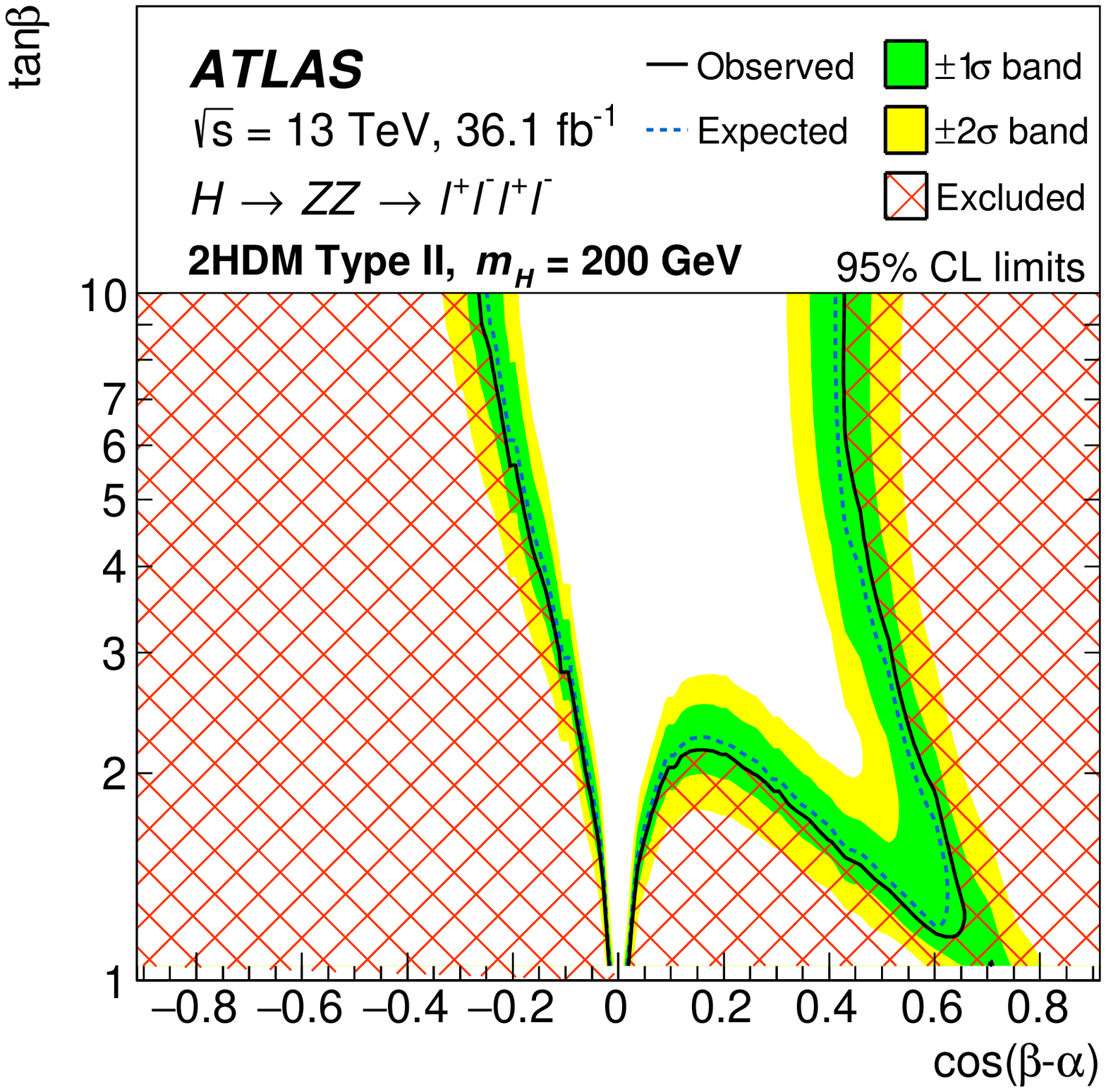}}
  \caption{The exclusion contour in the 2HDM~\subref{fig:2HDM_tanb_cba_mH200_sub1} Type-I and~\subref{fig:2HDM_tanb_cba_mH200_sub2} Type-II models for $\mH = 200$ \gev\ shown as a function of the parameters $\cos(\beta-\alpha)$ and $\tan\beta$. The green  and yellow  bands represent the $\pm$$1\sigma$ and $\pm$$2\sigma$ uncertainties in the expected limits. The hatched area shows the observed exclusion. 
  \label{fig:2HDM_tanb_cba_mH200}}
\end{figure}

\begin{figure}[tbp]
  \centering
  \subfigure[\label{fig:2HDM_tanb_mH_sub1}]{\includegraphics[width=0.49\textwidth]{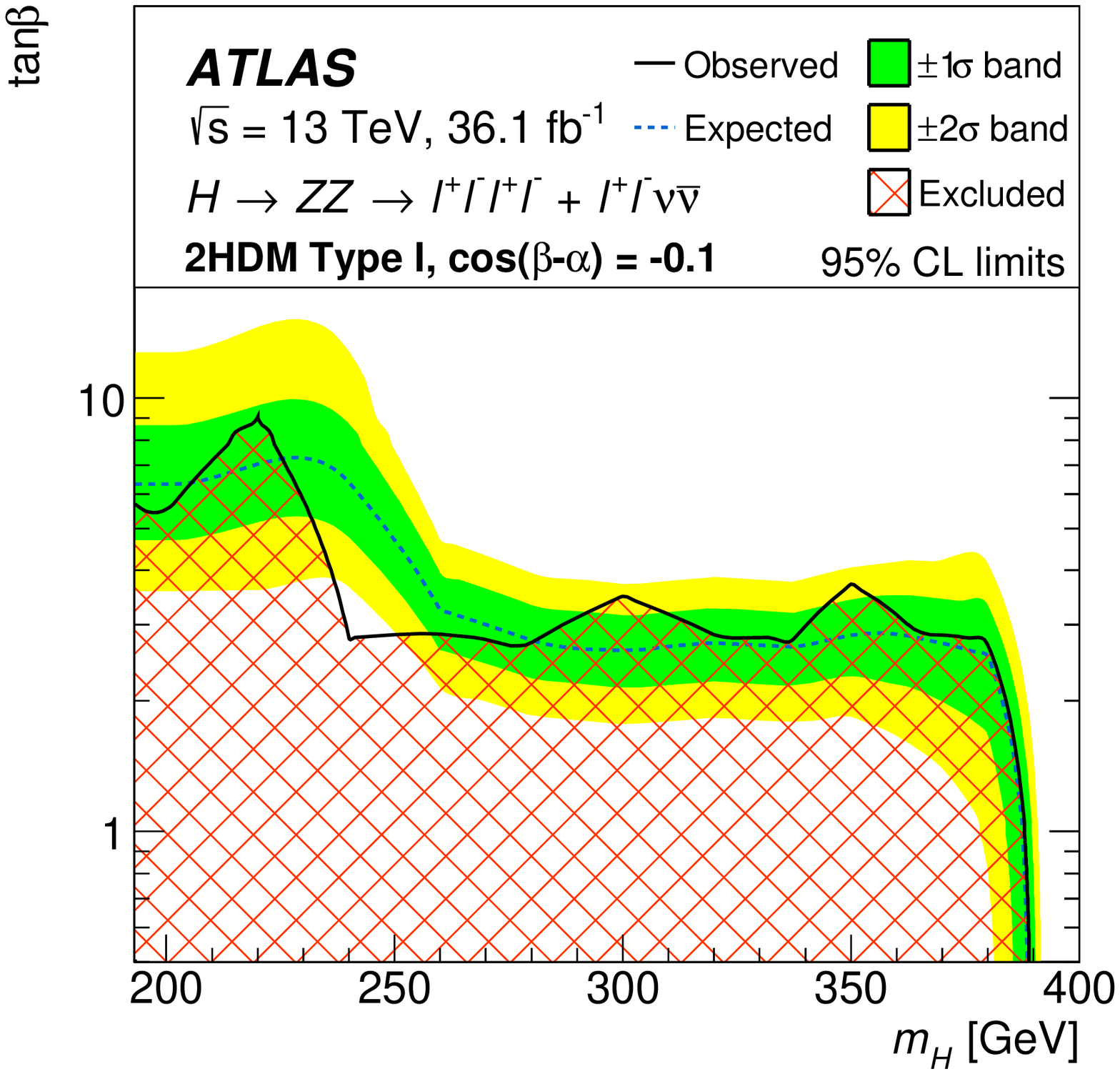}}
  \subfigure[\label{fig:2HDM_tanb_mH_sub2}]{\includegraphics[width=0.49\textwidth]{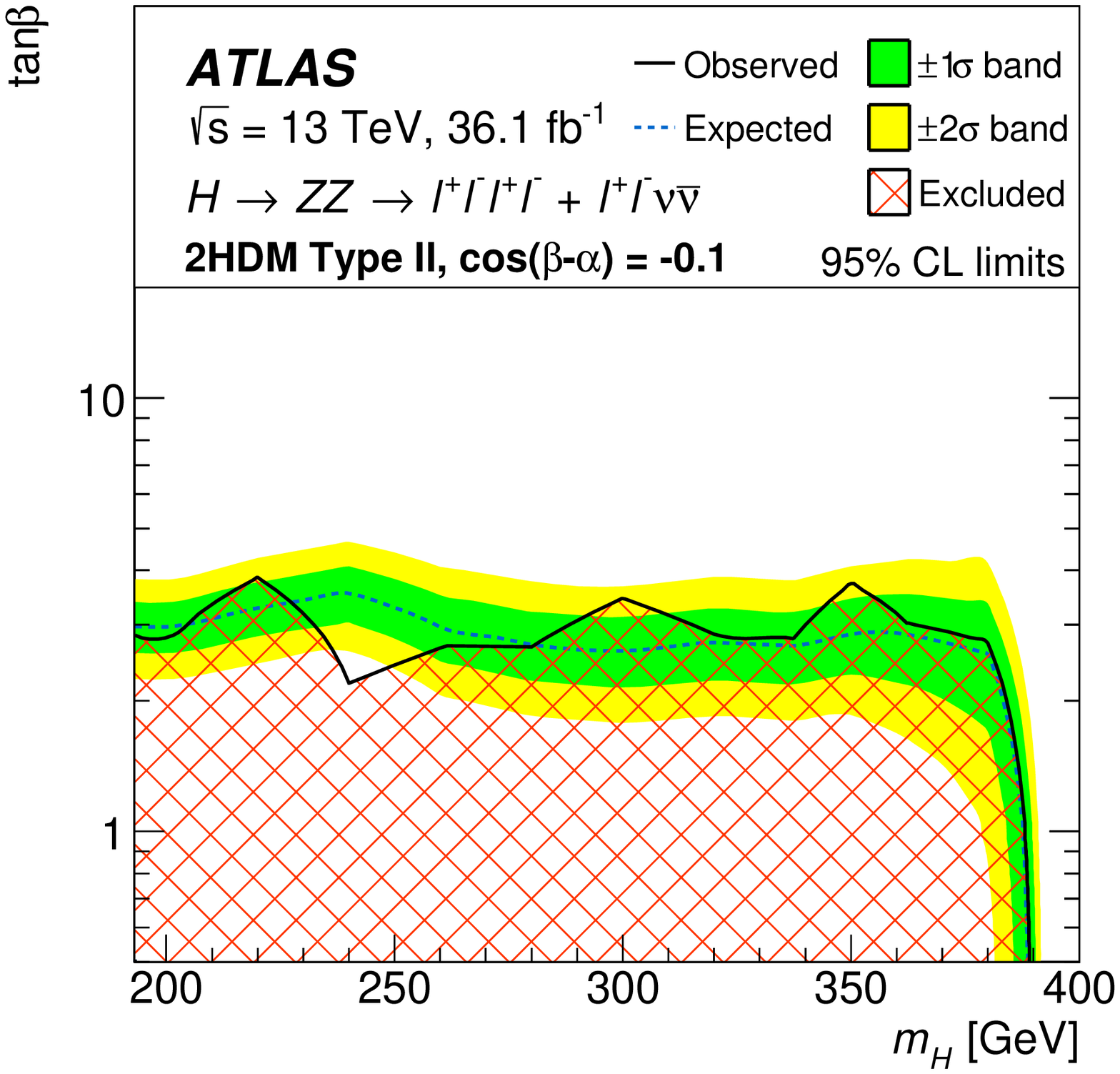}}
  \caption{The exclusion contour in the 2HDM \subref{fig:2HDM_tanb_mH_sub1} Type-I and \subref{fig:2HDM_tanb_mH_sub2} Type-II models for $\cos(\beta-\alpha)=-0.1$, shown as a function of the heavy scalar mass \mres\ and the parameter $\tan\beta$. The green  and yellow  bands represent the $\pm$$1\sigma$ and $\pm$$2\sigma$ uncertainties in the expected limits. The hatched area shows the observed exclusion.
  \label{fig:2HDM_tanb_mH}}
\end{figure}

Figure~\ref{fig:2HDM_tanb_cba_mH200} shows exclusion limits in the $\tan\beta$ versus $\cos(\beta-\alpha)$
plane for Type-I and Type-II 2HDMs, for a heavy Higgs boson with mass \mH =
200~\gev. This \mH value is chosen so that the assumption of a narrow Higgs boson is valid
over most of the parameter space, and the experimental sensitivity is maximal. At this low mass, 
only the \llll final state contributes to this result. The range of $\cos(\beta-\alpha)$ and
$\tan\beta$ explored is limited to the region where the assumption of a heavy narrow Higgs
boson with negligible interference is valid. When calculating the limits at a given choice of
$\cos(\beta-\alpha)$ and $\tan\beta$, the relative rates of ggF and VBF production in the fit are set
to the prediction of the 2HDM for that parameter choice. Figure~\ref{fig:2HDM_tanb_mH} shows
exclusion limits as a function of the heavy Higgs boson mass \mH and the parameter $\tan\beta$ for
$\cos(\beta-\alpha) = -0.1$. The white regions in the exclusion plots indicate regions of parameter
space which are not excluded by the present analysis. In these regions the cross section predicted
by the 2HDM is below the observed cross section limit. Compared with the results from 
\runa~\cite{HIGG-2013-20}, the exclusion presented here is almost twice as stringent.

\subsection{Spin-2 resonance interpretation}
\label{sec:res_Graviton}
\begin{figure}[tbp]
\centering
{\includegraphics[width=0.7\linewidth]{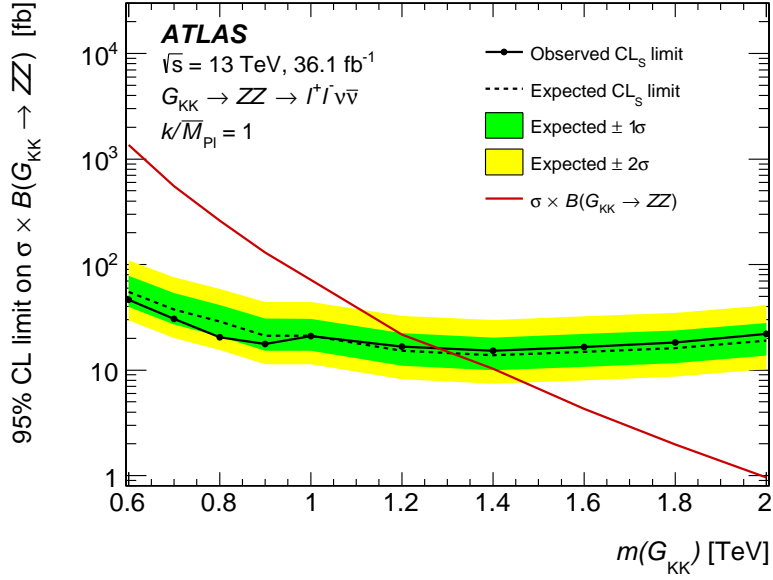}\label{fig:mtlim:graviton}}
    \caption{The upper limits at 95\% CL on cross section times branching ratio $\sigma \times\mbr(\Gratozz)$ for a KK graviton produced with $k/{\bar{M}_\mathrm{Pl}} = 1$. 
The green and yellow bands give the $\pm$$1\sigma$ and $\pm$$2\sigma$ uncertainties in the expected limits. 
The predicted production cross section times branching ratio as a function of the \grav mass \mgrav\ is shown by the red solid line.
}
\label{fig:graviton} 
\end{figure}

The results are also interpreted as a search for a Kaluza--Klein graviton excitation, \grav, in the context of the bulk RS model
using the \llvv final state because the \llll final state was found  to have negligible sensitivity for this type of model.
The limits on $\sigma \times\mbr(\Gratozz)$ at 95\% CL as a function of the KK graviton mass, \mgrav, are shown in Figure~\ref{fig:graviton} together with the predicted \grav\ cross section. A spin-2 graviton is excluded up to a mass of 1300~\gev. These limits have been extracted using the asymptotic approximation, and they were verified
 to be correct within about 4\% using pseudo-experiments. 

\FloatBarrier

\section{Summary}
\label{sec:summary}
A search is conducted for heavy resonances decaying into a pair of \Z bosons which subsequently decay
into \llll or \llvv final states. 
The search uses proton--proton collision data collected with the ATLAS detector during 2015 and 2016 at the Large Hadron
Collider at a centre-of-mass energy of 13~\tev\
corresponding to an integrated luminosity of \lumi. The results of the search are interpreted as
 upper limits on the production cross section of a spin-0 or spin-2 resonance. The mass range 
of the hypothetical resonances considered 
is between 200~\gev\ and 2000~\gev\ depending on the final state and the model considered. The spin-0 resonance is assumed to be a heavy scalar, whose dominant production modes are gluon--gluon fusion and vector-boson fusion and it is studied  in the narrow-width approximation and with the large-width assumption.  
In the case of the narrow-width approximation, limits on the production rate of a heavy scalar decaying into two $Z$ bosons 
are set separately for ggF and VBF production modes. 
Combining the two final states, 95\% CL upper limits range from 0.68~pb at $\mres = 242$~\gev\ to 11~fb at $\mres = 1200$~\gev\ 
for the gluon--gluon fusion production mode 
and from 0.41~pb at $\mres = 236$~\gev\ to 13~fb at $\mres = 1200$~\gev\ for the vector-boson fusion production mode.   
The results are also interpreted in the context of Type-I and Type-II two-Higgs-doublet
models, with exclusion contours given in the \tanb\ versus \cosab\ (for $\mH = 200$~\gev) and \tanb\ versus \mH\ planes.
This \mH value is chosen so that the assumption of a narrow Higgs boson is
valid over most of the parameter space and the experimental sensitivity is maximal.  The limits on
the production rate of a large-width scalar are obtained for widths of 1\%, 5\% and 10\% of the
mass of the resonance, with the interference between the heavy scalar and the SM Higgs boson as well
as the heavy scalar and the \ggZZ continuum taken into account.  In the framework of the Randall--Sundrum model 
with one warped extra dimension a graviton excitation spin-2 resonance 
with $\mgrav < 1300~\gev$ is excluded at 95\% CL.

\section*{Acknowledgements}

We thank CERN for the very successful operation of the LHC, as well as the
support staff from our institutions without whom ATLAS could not be
operated efficiently.

We acknowledge the support of ANPCyT, Argentina; YerPhI, Armenia; ARC, Australia; BMWFW and FWF, Austria; ANAS, Azerbaijan; SSTC, Belarus; CNPq and FAPESP, Brazil; NSERC, NRC and CFI, Canada; CERN; CONICYT, Chile; CAS, MOST and NSFC, China; COLCIENCIAS, Colombia; MSMT CR, MPO CR and VSC CR, Czech Republic; DNRF and DNSRC, Denmark; IN2P3-CNRS, CEA-DRF/IRFU, France; SRNSFG, Georgia; BMBF, HGF, and MPG, Germany; GSRT, Greece; RGC, Hong Kong SAR, China; ISF, I-CORE and Benoziyo Center, Israel; INFN, Italy; MEXT and JSPS, Japan; CNRST, Morocco; NWO, Netherlands; RCN, Norway; MNiSW and NCN, Poland; FCT, Portugal; MNE/IFA, Romania; MES of Russia and NRC KI, Russian Federation; JINR; MESTD, Serbia; MSSR, Slovakia; ARRS and MIZ\v{S}, Slovenia; DST/NRF, South Africa; MINECO, Spain; SRC and Wallenberg Foundation, Sweden; SERI, SNSF and Cantons of Bern and Geneva, Switzerland; MOST, Taiwan; TAEK, Turkey; STFC, United Kingdom; DOE and NSF, United States of America. In addition, individual groups and members have received support from BCKDF, the Canada Council, CANARIE, CRC, Compute Canada, FQRNT, and the Ontario Innovation Trust, Canada; EPLANET, ERC, ERDF, FP7, Horizon 2020 and Marie Sk{\l}odowska-Curie Actions, European Union; Investissements d'Avenir Labex and Idex, ANR, R{\'e}gion Auvergne and Fondation Partager le Savoir, France; DFG and AvH Foundation, Germany; Herakleitos, Thales and Aristeia programmes co-financed by EU-ESF and the Greek NSRF; BSF, GIF and Minerva, Israel; BRF, Norway; CERCA Programme Generalitat de Catalunya, Generalitat Valenciana, Spain; the Royal Society and Leverhulme Trust, United Kingdom.

The crucial computing support from all WLCG partners is acknowledged gratefully, in particular from CERN, the ATLAS Tier-1 facilities at TRIUMF (Canada), NDGF (Denmark, Norway, Sweden), CC-IN2P3 (France), KIT/GridKA (Germany), INFN-CNAF (Italy), NL-T1 (Netherlands), PIC (Spain), ASGC (Taiwan), RAL (UK) and BNL (USA), the Tier-2 facilities worldwide and large non-WLCG resource providers. Major contributors of computing resources are listed in Ref.~\cite{ATL-GEN-PUB-2016-002}.

\clearpage
\printbibliography
\clearpage
\begin{flushleft}
{\Large The ATLAS Collaboration}

\bigskip

M.~Aaboud$^\textrm{\scriptsize 137d}$,
G.~Aad$^\textrm{\scriptsize 88}$,
B.~Abbott$^\textrm{\scriptsize 115}$,
O.~Abdinov$^\textrm{\scriptsize 12}$$^{,*}$,
B.~Abeloos$^\textrm{\scriptsize 119}$,
S.H.~Abidi$^\textrm{\scriptsize 161}$,
O.S.~AbouZeid$^\textrm{\scriptsize 139}$,
N.L.~Abraham$^\textrm{\scriptsize 151}$,
H.~Abramowicz$^\textrm{\scriptsize 155}$,
H.~Abreu$^\textrm{\scriptsize 154}$
% ...
\bigskip
\\
% ...
$^{12}$ Institute of Physics, Azerbaijan Academy of Sciences, Baku, Azerbaijan\\
% ...
$^{88}$ CPPM, Aix-Marseille Universit{\'e} and CNRS/IN2P3, Marseille, France\\
% ...
$^{115}$ Homer L. Dodge Department of Physics and Astronomy, University of Oklahoma, Norman OK, United States of America\\
% ...
$^{119}$ LAL, Univ. Paris-Sud, CNRS/IN2P3, Universit{\'e} Paris-Saclay, Orsay, France\\
% ...
$^{137}$ $^{(a)}$ Facult{\'e} des Sciences Ain Chock, R{\'e}seau Universitaire de Physique des Hautes Energies - Universit{\'e} Hassan II, Casablanca; $^{(b)}$ Centre National de l'Energie des Sciences Techniques Nucleaires, Rabat; $^{(c)}$ Facult{\'e} des Sciences Semlalia, Universit{\'e} Cadi Ayyad, LPHEA-Marrakech; $^{(d)}$ Facult{\'e} des Sciences, Universit{\'e} Mohamed Premier and LPTPM, Oujda; $^{(e)}$ Facult{\'e} des sciences, Universit{\'e} Mohammed V, Rabat, Morocco\\
% ...
$^{139}$ Santa Cruz Institute for Particle Physics, University of California Santa Cruz, Santa Cruz CA, United States of America\\
% ...
$^{151}$ Department of Physics and Astronomy, University of Sussex, Brighton, United Kingdom\\
% ...
$^{154}$ Department of Physics, Technion: Israel Institute of Technology, Haifa, Israel\\
$^{155}$ Raymond and Beverly Sackler School of Physics and Astronomy, Tel Aviv University, Tel Aviv, Israel\\
% ...
$^{161}$ Department of Physics, University of Toronto, Toronto ON, Canada\\
% ...
$^{a}$ Also at Department of Physics, King's College London, London, United Kingdom\\
$^{b}$ Also at Institute of Physics, Azerbaijan Academy of Sciences, Baku, Azerbaijan\\
$^{c}$ Also at Novosibirsk State University, Novosibirsk, Russia\\
$^{d}$ Also at TRIUMF, Vancouver BC, Canada\\
$^{e}$ Also at Department of Physics {\&} Astronomy, University of Louisville, Louisville, KY, United States of America\\
% ...
$^{*}$ Deceased
\end{flushleft}

\end{document}